\documentclass[article,useAMS,usenatbib]{mn2e}

\usepackage{amsmath}
\usepackage{graphicx} 

\usepackage{flafter}
\usepackage{float}
\usepackage{fixltx2e}

\newcommand{\ddp}[2]{\frac{\upartial #1}{\upartial #2}}
\newcommand{\dddp}[2]{\dfrac{\upartial #1}{\upartial #2}}
\newcommand{\lddp}[2]{\upartial #1/\upartial #2}

\newcommand{\bn}{\boldsymbol{n}}
\newcommand{\bom}{\boldsymbol{\omega}}
\newcommand{\eexp}{\mathrm{e}}
\newcommand{\bJ}{\boldsymbol{J}}
\newcommand{\bth}{\boldsymbol{\theta}}

\newcommand{\bx}{\boldsymbol{x}}
\newcommand{\bv}{\boldsymbol{v}}

\newcommand{\sech}{\mathrm{sech}}

\newcommand{\Rd}{h_{\mathrm{R}}}
\newcommand{\Rs}{R_{\mathrm{s}}}

\newcommand{\Omp}{\Omega_\mathrm{p}}
\newcommand{\omp}{\omega_\mathrm{p}}

\newcommand{\Kpc}{~\mathrm{kpc}}

\newcommand{\kmsec}{~\mathrm{km}~\mathrm{s}^{-1}}
\newcommand{\kmsect}{~\mathrm{km^2}~\mathrm{s}^{-2}}
\newcommand{\kmseckpc}{~\mathrm{km}~\mathrm{s}^{-1}~\mathrm{kpc}^{-1}}

\newcommand{\vc}{v_{\mathrm{c}}}

\newcommand{\de}{\mathrm{d}}

\newcommand{\Rg}{R_\mathrm{g}}

\newcommand{\RNum}[1]{\uppercase\expandafter{\romannumeral #1\relax}}

\newcommand{\Ec}{E_\mathrm{c}}

\newcommand{\ldd}[2]{\de #1/\de #2}
\newcommand{\dd}[2]{\frac{\de #1}{\de #2}}
\newcommand{\pare}[1]{\left(#1\right)}
\newcommand{\paresq}[1]{\left[#1\right]}
\newcommand{\parec}[1]{\left\{#1\right\}}
\newcommand{\av}[1]{\protect\langle #1 \rangle}
\newcommand{\avvR}{\av{v_R}}

\newcommand{\avvz}{\av{v_z}}
\newcommand{\img}{\mathrm{i}}
\newcommand{\Rep}{\operatorname{Re}}
\newcommand{\Eq}[1]{Eq.~(\ref{#1})}
\newcommand{\Eqs}[2]{Eqs.~(\ref{#1})-(\ref{#2})}

\newcommand{\Fig}[1]{Fig.~\ref{#1}}

\newcommand{\Phia}{\Phi_\mathrm{a}}

\newcommand{\uza}{u_z^\mathrm{a}}
\newcommand{\dvz}{\Delta \av{v_z}}

\newcommand{\uRa}{u_R^\mathrm{a}}

\newcommand{\Sec}[1]{Section~\ref{#1}}

\newcommand{\uzo}{u_{z,1}}

\newcommand{\uRo}{u_{R,1}}

\newcommand{\calF}{{\cal F}}

\newcommand{\hs}{h_{\mathrm{s}}}

\newcommand{\phis}{\phi_{\mathrm{s}}}

\newcommand{\sgn}{\mathrm{sgn}}
\newcommand{\sigmaR}{\tilde{\sigma}_R}
\newcommand{\sigmaz}{\tilde{\sigma}_z}

\newcommand{\hS}{\hat{\Sigma}}
\newcommand{\hphi}{\hat{\phi}}

\title[Perturbed distribution functions]{Modelling the Galactic disc: perturbed distribution functions in the presence of spiral arms} \author[G. Monari et al.]  {
  Giacomo~Monari\thanks{Email:~\texttt{giacomo.monari@astro.unistra.fr}},  Benoit~Famaey, Arnaud~Siebert\\ Observatoire astronomique de
  Strasbourg, Universit\'e de Strasbourg, CNRS UMR 7550, 11 rue de
  l'Universit\'e, 67000 Strasbourg, France } \date{Released 2015 Xxxxx
  XX}

\pagerange{\pageref{firstpage}--\pageref{lastpage}} \pubyear{2015}

\def\LaTeX{L\kern-.36em\raise.3ex\hbox{a}\kern-.15em
    T\kern-.1667em\lower.7ex\hbox{E}\kern-.125emX}

\begin{document}

\label{firstpage}

\maketitle

\begin{abstract}
Starting from an axisymmetric equilibrium distribution function (DF)
in action space, representing a Milky Way thin disc stellar
population, we use the linearized Boltzmann equation to explicitly
compute the response to a three-dimensional spiral potential in terms
of the perturbed DF. This DF, valid away from the main resonances,
allows us to investigate a snapshot of the velocity distribution at
any given point in three-dimensional configuration space. Moreover,
the first order moments of the DF give rise to non-zero radial and
vertical bulk flows -- namely breathing modes -- qualitatively similar
to those recently observed in the extended Solar neighbourhood. We
show that these analytically predicted mean stellar motions are in
agreement with the outcome of test-particle simulations. Moreover, we
estimate for the first time the reduction factor for the vertical bulk
motions of a stellar population compared to the case of a cold
fluid. Such an explicit expression for the full perturbed DF of a thin
disc stellar population in the presence of spiral arms will be helpful
in order to dynamically interpret the detailed information on the
Milky Way disc stellar kinematics that will be provided by upcoming
large astrometric and spectroscopic surveys of the Galaxy.
\end{abstract}


\section{Introduction}

The primary objective of the current and future large spectroscopic
and astrometric surveys of the Milky Way, culminating with the Gaia
mission \citep{Prusti}, will be to provide a detailed dynamical model
of the Galaxy, including all of its components, and giving us insight
into its structure, its formation and its evolutionary history.

The top-down dynamical approach consists in producing {\it ab initio}
simulations of Milky Way-like galaxies in a cosmological context. This
approach can be useful to understand some general features of galaxy
formation \citep[e.g.,][]{Minchev2014}. However, it is not flexible
enough to produce an acceptable model for the wide range of extremely
detailed data soon to be available for our own Galaxy. On the other
side, the bottom-up approach for dynamical modelling consists in
starting from the actual Galactic data, rather than from simulations,
in order to construct a model of the Galaxy. To avoid the redundancy
and computational waste of representing the orbits of every single
particle in the model, one can use a phase-space distribution function
(DF) to represent each population of constituent particles
\citep[typically, various stellar populations and dark matter, see
  e.g.][]{Piffl2015,BinneyPiffl}. The model-building generally starts
from the assumptions of dynamical equilibrium and axisymmetry. These
assumptions allow us to make use of Jeans' theorem constraining the DF
to depend only on three integrals of motion, which can typically be
chosen to be the radial, azimuthal, and vertical action
variables. However, one should remember, especially when modelling the
stellar populations of the Galactic disc, that the Galaxy is obviously
not axisymmetric, as it harbours a central bar as well as spiral
arms. Such perturbations can obviously be treated through perturbation
theory, whose foundations in the case of flat 2D discs have been laid
down by \citet{Kalnajs1971}. For instance, following up on the work of
\citet{BinneyLacey} who derived the orbit-averaged Fokker-Planck
equation for a 2D stellar disc, recent investigations
\citep[e.g.,][]{Fouvry} have focused on the long-term secular
evolution of such a flat disc by means of diffusion through
action-space at resonances, producing ridges in action-space. Here, we
are rather interested in the present-day perturbed distribution
function in the action-angle space of the unperturbed Hamiltonian, in
the presence of a 3D spiral arm perturber, which could be fitted to a
snapshot of the Galaxy taken by current and upcoming large
surveys. Our philosophy is thus closer to that of
\citet{McMillan2013}, except that the shape of the perturbed DF will
be computed directly from the linearized Boltzmann equation. Moreover,
in this paper, we will first concentrate only on the response {\it
  away} from the main resonances, the extremely interesting effects
expected at resonances, as well as the effect of resonance overlaps of
multiple perturbers \citep[e.g.][]{Quillen,MinchevFamaey}, being the
subject of further analytical work.

One potential issue with assuming axisymmetry in order to produce a
benchmark model of the Galaxy from a Galactic survey snapshot is that
it is not clear that the fundamental parameters entering the model,
such as the peculiar motion of the Sun, will not be biased by forcing
the model to fit observed non-axisymmetric features that are not
present in the axisymmetric model itself. This could for instance
explain why current determinations of the peculiar motion of the Sun
are discrepant with each other when using local or non-local tracers
\citep{Schonrich2010,Schonrich2012}. It would thus be extremely
useful, especially when modelling the Galactic disc stellar
populations, to directly include in the model the response of the
stellar DF to the bar or to spiral arms.

In this contribution, we make a step in this direction by analytically
investigating the response of a typical stellar population
representative of the thin disc of the Galaxy to a 3D perturbing
spiral potential. More specifically, we are able to provide the fully
explicit form of the perturbed DF in angle-action variables, which
could later on be used for dynamical modelling of the disc stellar
populations. However, the main problem with including spirals in our
model is that the nature and origin of spiral arms in galactic discs
are still mostly unknown. Recent numerical investigations indicate
that spirals might consist of multiple long-lived ($\sim 10$ galaxy
rotations) modes \citep{SellwoodCarlberg2014}, which do not appear to
be strictly static as in the classical density wave picture, but are
nevertheless genuine standing wave oscillations with fixed shape and
pattern speed. The response to these waves away from the main
resonances and the regions where nonlinear coupling between the modes
is important can then be computed from perturbation theory, and can in
principle be linearly added to each other. Hence, it is interesting to
consider the response of the DF to a single such mode, which we will
assume here to have non-varying amplitude but which could be later
generalized to varying amplitudes too. In the present work, we will
concentrate on the response of a given disc stellar population in
equilibrium to a perturbing spiral potential in 3D, but we do not
investigate yet the conditions for self-consistency. 

In \Sec{sect:DF}, we recall the basics of action-angle variables and
equilibrium distributions functions in action space. The response of a
stellar population, represented by a given equilibrium distribution,
to a perturbing potential is then presented in \Sec{sect:lCBE}, by
means of the linearized collisionless Boltzmann equation. We compute
both the perturbed DF and its first order moments, giving the mean
stellar motions. The results for the specific case of a spiral
perturber are presented in \Sec{sect:results}, and conclusions are
drawn in \Sec{sect:conclusions}.

\section{Equilibrium distribution functions in action space}\label{sect:DF}

It is well-known that, in realistic axisymmetric and time-independent
Galactic potentials, most orbits are regular (i.e., they are
quasiperiodic in the sense that their Fourier transforms have only
discrete frequencies that are integer linear combinations of 3
fundamental frequencies) and thus have three isolating integrals of
the motion. Each triplet of them specifies a particular orbit in the
potential of the Galaxy. Jeans' theorem then tells us that the
equilibrium stellar phase-space DF of any component of the Galaxy,
$f_0$, should depend only on these three integrals, which makes $f_0$
automatically a solution of the collisionless Boltzmann equation:
\begin{equation}\label{eq:CBE}
\frac{{\rm d}f_0}{{\rm d}t} = 0.
\end{equation}
There are in principle an infinity of sets of isolating integrals of
the motion to choose from.

On the other hand, if one of the configuration space variables of a
dynamical system is absent from the Hamiltonian, then its conjugate
momentum is itself an integral of the motion, as is evident from
Hamilton's equations. Conversely, if an integral of the motion has a
canonically conjugate variable, the Hamiltonian does not depend on
that variable. Hence by choosing three isolating integrals of the
motion having canonically conjugate variables, the Hamiltonian can be
written in its simplest form, purely as a function of the three
integrals of the motion. This makes such a choice of integrals
particularly appealing. Such integrals are called the ``action
variables" $\bJ$, and correspond to new generalized momenta. Their
canonically conjugate variables are called the ``angle variables"
$\bth$, because they can be normalized such that the position in
phase-space is $2\upi$-periodic in them.

The equations of motion (Hamilton's equations) are conveniently
expressed as
\begin{equation}\label{eq:mot0}
  \dot{\bth}=\ddp{H_0}{\bJ}=\bom(\bJ),\quad \dot{\bJ}=-\ddp{H_0}{\bth}=0. 
\end{equation}
For a star in an axisymmetric disc galaxy, for which the usual
phase-space coordinates are the cylindrical coordinates $(R,\phi,z)$
and their associated velocities $(v_R,v_\phi,v_z) \equiv
(\dot{R},R\dot{\phi},\dot{z})$, $\bJ=(J_R,J_\phi,J_z)$ are the
actions, $\bth=(\theta_R,\theta_\phi,\theta_z)$ the angles, and
$H_0(\bJ)$ is the Hamiltonian corresponding to the axisymmetric
time-independent potential $\Phi_0$.  The motion is as simple as one
can imagine, since the actions $\bJ$ are constant in time, and define
orbital tori on which the angles just evolve linearly with time, i.e.,
$\bth(t)=\bom t+\bth_0$, where $\bom(\bJ)\equiv\lddp{H_0}{\bJ}$ are
the orbital fundamental frequencies.

One of the drawbacks is that we can write analytical relations between
the action-angles $(\bJ,\bth)$ and the usual phase-space coordinates
$(\bx,\bv)$ only in some rare cases\footnote{Note that, for any choice
  of integrals, the third integral cannot, in general, be written
  analytically for a disc galaxy, apart when the vertical motion is
  considered separable from the horizontal one as assumed here, or if
  the potential is of St\"ackel form
  \citep[e.g.,][]{FamaeyDejonghe2003}. \citet{Bienayme2015} provide
  typical analytic approximations for the third integral in more
  realistic potentials, based on the St\"ackel approximation, but the
  corresponding actions are not analytic either.} of potentials
$\Phi_0$. 
But the advantages are numerous. First of all, in an equilibrium
configuration for the Galaxy, the phase of the stars, $\bth$, are
uniformly distributed (phase mixed) on orbital tori specified by $\bJ$
alone, and the phase-space density of stars $f_0(\bJ) \de^3\bJ$ is
just the number of stars $dN$ in a given infinitesimal action range
divided by a factor $(2\upi)^3$. Secondly, the actions are
adiabatically invariant for a slow secular evolution of the Galactic
potential. And finally, they are very natural coordinates for
perturbation theory: the linearized collisionless Boltzmann equation
takes a rather simple form with these variables (see \Sec{sect:lCBE}).

For simplicity, we are going to work here in the epicyclic and
adiabatic approximations \citep[for various more rigorous ways of
  evaluating the actions, see,
  e.g.,][]{McGill1990,McMillan2008,BinneyMcM2011,Binney2012,Bovy2013,Sanders2014},
assuming separable motion in the vertical and horizontal
directions. The epicyclic approximation is roughly valid for the thin
disc we want to model here, i.e. for not too eccentric orbits and
close enough to the Galactic plane. It consists in locally
approximating the radial and vertical motions of an orbit of angular
momentum $L_z\equiv Rv_\phi$ with harmonic motions, i.e., with an
effective potential in the meridional plane of the form
\begin{equation}
  \Phi_{0, {\rm eff}} = \Phi_0 + \frac{L_z^2}{2R^2}\simeq \Ec + \Phi_{0,R} + \Phi_{0,z},
\end{equation}
where $\Phi_{0,R}\equiv \kappa^2\pare{R-\Rg}^2/2$, $\Phi_{0,z}\equiv
\nu^2 z^2/2$, and the radial and vertical epicyclic frequencies,
$\kappa$ and $\nu$, are evaluated at $\Rg$, the radius of a circular
orbit of angular momentum $L_z$, whose energy is $\Ec$. The techniques
and results developed in this paper are nevertheless in principle generalizable to
more precise and general estimates of the actions for a wider range of
orbits, which will be the topic of further papers. Within the
adiabatic and epicyclic approximations, the actions $(J_R,J_\phi,J_z)$
are approximated by the following explicit analytic form:
\begin{align}\label{eq:transfJ}
  J_\phi &= \frac{1}{2 \upi} \int_{0}^{2 \upi} {\rm d}\phi L_z = L_z, \nonumber \\ 
  J_z &\simeq \frac{1}{\upi} \int_{z_{\rm min}}^{z_{\rm
      max}} {\rm d}z \sqrt{ 2 [E_z - \Phi_{0,z}] } = \frac{E_z}{\nu},\\ 
  J_R & \simeq \frac{1}{\upi} \int_{R_{\rm min}}^{R_{\rm max}} {\rm d}R
  \sqrt{2 (E_R - \Phi_{0,R}) } = \frac{E_R}{\kappa}
  \nonumber,
\end{align} 
where $E_R=v_R^2/2+\kappa^2(R-\Rg)^2/2$ is the
radial epicyclic energy and $E_z=v_z^2/2+\nu^2z^2/2$ is the vertical
energy. The canonically conjugate angle
variables can then also be expressed explicitly
\citep{Dehnen1999,BT2008} as:
\begin{align}\label{eq:transftheta}
  \theta_\phi &\simeq \phi+\Delta\phi, \nonumber \\
  \theta_z &\simeq \tan^{-1} \pare{-\frac{v_z}{\nu z}}, \\ 
  \theta_R &\simeq \tan^{-1}\pare{-\frac{v_R}{\kappa (R-\Rg)}},
\end{align}
where\footnote{Rigorously speaking $\theta_\phi$, the canonical
  conjugate of $J_\phi$, should also include a term dependent on the
  vertical motion $-J_z(\de\ln\nu/\de J_\phi)\cos\theta_z\sin\theta_z$
  which in typical thin disc situations is tiny, much smaller than the
  already small $-J_R(\de\ln\kappa/\de J_\phi)\sin(2\theta_R)$, and
  which therefore we omit.}
\begin{equation}
  \Delta \phi\equiv -\frac{\gamma}{\Rg}\sqrt{\frac{2 J_R}{\kappa}}\sin\theta_R-
                    \frac{J_R}{2}\dd{\ln \kappa}{J_\phi}\sin(2\theta_R),
\end{equation}
with
\begin{equation}\label{gamma}
\gamma \equiv 2 \Omega/\kappa, 
\end{equation}
and $\Omega$ the angular circular frequency evaluated at
$\Rg(J_\phi)$. Finally, the orbital frequencies are approximated by
\begin{align}\label{eq:transffreq}
  \omega_\phi &\simeq \Omega + (\ldd{\kappa}{J_\phi})J_R, \nonumber \\
  \omega_z &\simeq \nu, \\
  \omega_R &\simeq \kappa. \nonumber
\end{align}

The possible choices of realistic DFs to represent the different
components of the Galactic disc are again numerous \citep[see
  e.g.,][]{Binney2010,Binney2014}. Here we will make the simplest
assumption, i.e. that the axisymmetric thin disc is well represented
by a Schwarzschild DF \citep{BT2008}, i.e.,
\begin{equation}\label{schwarzschild}
  f_0(J_R,J_\phi,J_z)=\frac{\gamma \tilde{\Sigma}_0 {\rm exp}(-{\Rg}/{\Rd})}
  {4\pare{2\upi}^{3/2}\sigmaR^2\sigmaz z_0} {\rm exp} \left(- \frac{J_R\kappa}{\sigmaR^2} - \frac{J_z\nu}{\sigmaz^2} \right),
\end{equation}
where $\sigmaR$, $\sigmaz$, $\kappa$, $\nu$, and $\gamma$ are all
functions of $J_\phi$ through a chosen dependence on
$\Rg(J_\phi)$. Note however that most results in \Sec{sect:lCBE} will
be fully independent of this particular choice for $f_0$.

\section{Linearized collisionless Boltzmann equation}\label{sect:lCBE}
\subsection{General solution}\label{sect:lCBE_gen}
In this section, we will consider a small perturbation to the
potential, denoted $\epsilon\Phi_1$ where $\epsilon \ll 1$, $\Phi_1$
has the same order of magnitude as the axisymmetric background
potential $\Phi_0$, and the total potential is
$\Phi=\Phi_0+\epsilon\Phi_1$. Instead of searching for new
action-angle variables for the perturbed Hamiltonian $H_1 = H_0 +
\epsilon \Phi_1$, we will continue here to work with the variables
defined within the {\it unperturbed} Hamiltonian $H_0$. These are
obviously no longer action-angle variables within $H_1$, but they
remain canonical. The following calculations in this
\Sec{sect:lCBE_gen} are fully independent from the specific
action-angle estimate and choice of DF mentioned at the end of
\Sec{sect:DF}. We will move to specific predictions involving our
specific choice of variables only in \Sec{sect:lCBE_epi}.

With such a perturbation, the DF becomes, to first order in
$\epsilon$, $f=f_0+\epsilon f_1$, which should still be a solution of
the collisionless Boltzmann equation, \Eq{eq:CBE}. To first order in
$\epsilon$ (i.e. dropping higher-order terms), this leads to the {\it
  linearized} collisionless Boltzmann equation, which reads
\citep[Eqs. 5.13 \& 5.14 of][]{BT2008}:
\begin{equation}
\label{LCBE}
\frac{{\rm d}f_1}{{\rm d}t} + \paresq{f_0,\Phi_1} = 0,
\end{equation}
where the time-derivative of $f_1$ is a {\it total} derivative and
$\paresq{f_0,\Phi_1}$ is the Poisson bracket estimated along the
unperturbed orbits. It thus appears immediately that for a given
axisymmetric equilibrium DF, $f_0$, and a given perturbing potential,
$\Phi_1$, the response $f_1$ can be computed.

Integrating \Eq{LCBE} within angle-action coordinates leads to
\begin{equation}
\label{ILCBE}
  f_1(\bJ,\bth,t)=\int_{-\infty}^t\de
  t'\ddp{f_0}{\bJ'}(\bJ')\cdot\ddp{\Phi_1}{\bth'}(\bJ',\bth',t'),
\end{equation}
where the coordinates $(\bJ',\bth')$ correspond to the orbits in the
unperturbed potential. Note that the perturbing potential $\Phi_1$ is
assumed to have an explicit dependence on time.

Since the angle variables are defined such that the position in
phase-space is $2\upi$-periodic in them, we consider only cases where
$\Phi_1$ is cyclic in the angle coordinates, i.e.,
\begin{equation}\label{eq:phi1}
  \Phi_1\lvert_{\theta_i}=\Phi_1\lvert_{\theta_i+2\upi},
\end{equation}
where $\theta_i$ is any of the angle coordinates and the vertical line
means that $\Phi_1$ is evaluated keeping constant all the other
variables. Then, $\Phi_1$ can be expanded in a Fourier series as
\begin{equation}\label{eq:Phi1}
  \Phi_1(\bJ,\bth,t)= \Rep\parec{{\cal G}(t)\sum_{\bn} c_{\bn}(\bJ) \eexp^{\img \bn \cdot \bth}},
\end{equation}
where ${\cal G}(t)$ controls the strength of the perturbation as a
function of time. It is convenient to factorize this function into two
factors, ${\cal G}(t) = g(t) h(t)$, where $g(t)$ is a well behaved
function controlling the general amplitude of the perturbation, and
$h(t)$ is a periodic sinusoidal function of frequency $\omp$, which
can account for a perturbing potential rotating with a fixed pattern
speed. Hereabove, $\bn$ is a triple of indexes running from $-\infty$
to $\infty$.  Then \Eq{ILCBE} becomes
\begin{align}\label{eq:int}
  f_1(\bJ,\bth,t) =&\Rep\Bigg\{i\ddp{f_0}{\bJ}(\bJ)\cdot\sum_{\bn}\bn c_{\bn}(\bJ) \nonumber \\
                   &\quad \times \int_{-\infty}^t\de t'g(t')h(t')\eexp^{\img \bn \cdot \bth'(t')}\Bigg\}.
\end{align}
Integrating by parts, the solution of the integral in \Eq{eq:int} is
\begin{align}\label{eq:f1solve}
  \int_{-\infty}^t\de t'&g(t')h(t')\eexp^{\img \bn \cdot \bth'(t')}=\nonumber \\ &\sum_{k=0}^\infty
  \paresq{(-1)^k\frac{h(t')\eexp^{\img\bn\cdot\bth'(t')}g^{(k)}(t')}{(\img\bn\cdot\bom+\img\omp)^{k+1}}}_{-\infty}^t.
\end{align}
We assume that the perturbation and its time derivatives are null far
back in time, i.e., $g^{(k)}(-\infty)=0$. Moreover, we assume in the
following that the amplitude of the perturbation is constant at the
present time $t$, hence $g^{(0)}(t)=1$, and $g^{(k)}(t)=0$, for
$k=1,...,\infty$. This finally leads to
\begin{equation}\label{eq:f1}
  f_1(\bJ,\bth,t)=\Rep\parec{\ddp{f_0}{\bJ}(\bJ)\cdot\sum_{\bn}\bn c_{\bn}(\bJ)
  \frac{h(t)\eexp^{\img\bn\cdot\bth}}{\bn\cdot\bom+\omp}}.
\end{equation}
Within the above assumption of a currently non-varying amplitude of
the perturbation, this solution is completely general and independent
of any choice of action-angle coordinates and of any choice of a
particular form of the axisymmetric equilibrium DF $f_0$. Note that
\cite{CarlbergSellwood1985} and \cite{Carlberg1987} have on their side
investigated the lasting changes in the distribution function after a
transient spiral has come and gone. While similar in spirit to the
present work, the goal was very different and needed to consider the
second-order response, since to first-order, after the spiral has
vanished, the DF goes back to its initial state through
phase-mixing. Our approach is rather approximating what happens when
the amplitude of the spiral wave reaches a plateau at its maximum.

\subsection{Fourier modes perturbing potential within the epicyclic
  approximation}\label{sect:lCBE_epi}
To be more specific, we now consider a perturbing potential of the
form
\begin{equation}
  \Phi_1(R,\phi,z,t)=\Rep\parec{\Phia(R,z)\eexp^{\img m(\phi-\Omp t)}},
\end{equation} 
i.e., a pure Fourier mode in $\phi$, which is a good approximation for
the potential of a given spiral arm mode, or the bar (at least away
from the center of the Galaxy). Note that we only consider hereafter
plane-symmetric potentials $\Phia(R,|z|)$, thereby not addressing
perturbations such as corrugation waves.  Here, $\Omp$ is simply the
pattern speed, while $m$ is the azimuthal wavenumber (e.g., $m=2$ for
the bar or a 2-armed spiral, $m=4$ for a 4-armed spiral).

At this point, in order to specify the above solution $f_1$
(Eq.~\ref{eq:f1}) within that perturbing potential, we have to rewrite
$\Phi_1$ as in \Eq{eq:Phi1}. To do so, we approximate $\Phia(R,z)$
close to the plane as
\begin{equation}
  \Phia(R,z)\approx \Phia(R,0) + \frac{1}{2}\ddp{^2\Phia}{z^2}(R,0)z^2,
\end{equation}
which is valid in the same range of $z$ as the epicyclic
approximation. So, $\Phi_1$ becomes
\begin{equation}
  \Phi_1\approx \Phi_{1,R}(R,\phi)+\Phi_{1,z}(R,\phi,z),
\end{equation}
where
\begin{align}
 \Phi_{1,R}&\equiv\Rep\parec{\Phia(R,0)\eexp^{\img m(\phi-\Omp
    t)}}, \nonumber \\
 \Phi_{1,z}&\equiv\Rep\parec{\ddp{^2\Phia(R,0)}{z^2}\frac{z^2}{2}\eexp^{\img m(\phi-\Omp
    t)}}.
\end{align}
We start with $\Phi_{1,R}$. The radial motion in the
epicyclic approximation is written
\begin{equation}
  R=\Rg\pare{1-e\cos\theta_R},
\end{equation}
where
\begin{equation}
e(J_R,J_\phi)\equiv\sqrt{2J_R/(\kappa\Rg^2)} 
\end{equation}
is the eccentricity of the orbit. We consider orbits with low $e$, for
which the epicyclic approximation holds. Using the definition of $e$
and the mapping of \Eq{eq:transfJ} and \Eq{eq:transftheta}, we can
rewrite $\Phi_{1,R}$ and expand it in powers of $e$, dropping all the
terms that are $O(e^2)$, to obtain \citep[e.g.,][]{Weinberg1994}
\begin{align}
  \Phi_{1,R}&=\Rep\parec{\Phia(R,0)\eexp^{\img m (\theta_\phi-\Delta\phi-\Omp t)}} \nonumber \\
  &\approx \Rep\Bigg\{\Bigg[ \Phia(\Rg,0)(1+\img m e\gamma\sin\theta_R)+ \nonumber \\
  &\quad -\ddp{\Phia}{R}(\Rg,0)e\cos\theta_R \Bigg] \eexp^{\img m (\theta_\phi-\Omp t)}\Bigg\},
\end{align}
where $\gamma$ is defined as in \Eq{gamma}. Note that the function
$h(t)$ in \Eq{eq:Phi1} is just $h(t)={\rm exp}(- \img m \Omega_p t)$
in this case, and the frequency $\omp$ in \Eqs{eq:f1solve}{eq:f1} is
thus $\omp=-m\Omp$. We can now evaluate the Fourier coefficients for
$\Phi_{1,R}$ in the traditional way
\begin{align}\label{eq:cR}
  &c_{j k l}^R(J_R,J_\phi,J_z)=\frac{1}{(2\upi)^3}
  \int_{0}^{2\upi}\de\theta_R \int_{0}^{2\upi}\de\theta_\phi \nonumber \\  &\qquad \times \int_{0}^{2\upi}\de\theta_z \Phia(R,0)\eexp^{\img m (\theta_\phi-\Delta\phi)}\eexp^{-\img(j\theta_R+k\theta_\phi+l\theta_z)} \nonumber \\
  &\qquad\approx\delta_{k m} \delta_{l 0} \Bigg\{\Bigg[\delta_{j 0}+\delta_{|j|1}\frac{k}{2}\sgn(j)
    \gamma e\Bigg]\Phia(\Rg,0) \nonumber \\
  &\qquad\quad -\delta_{|j|1}\frac{\Rg}{2}e\ddp{\Phia}{R}(\Rg,0)\Bigg\},
\end{align}
where $\delta$ is the Kronecker delta. We can now also treat
$\Phi_{1,z}$ in the same way, replacing $\Phia(R,0)$ herabove by
$\ddp{^2\Phia(R,0)}{z^2}\frac{z^2}{2}$. From \Eq{eq:transftheta}, we
note that $z^2$ can be expressed as
\begin{equation}
  z^2=\frac{2J_z}{\nu}\cos^2\theta_z=\frac{J_z}{\nu}\sum_{l=-1}^1\frac{\eexp^{\img 2l\theta_z}}{2^{|l|}}.
\end{equation}
The Fourier coefficients for $\Phi_{1,z}$ are then
\begin{align}\label{eq:cz}
  &c_{j k l}^z(J_R,J_\phi,J_z)=\frac{1}{(2\upi)^3}
  \int_{0}^{2\upi}\de\theta_R \int_{0}^{2\upi}\de\theta_\phi \nonumber \\  &\qquad \times \int_{0}^{2\upi}\de\theta_z \ddp{^2\Phia(R,0)}{z^2}\frac{z^2}{2}\eexp^{\img m (\theta_\phi-\Delta\phi)}\eexp^{-\img(j\theta_R+k\theta_\phi+l\theta_z)} \nonumber \\
  &\qquad\approx\frac{1}{2}\delta_{k m} 
  \pare{\delta_{l 0}+\frac{\delta_{|l| 2}}{2}}\frac{J_z}{\nu}\Bigg\{\Bigg[\delta_{j 0}+\delta_{|j|1}
    \frac{k}{2}\sgn(j)\gamma e\Bigg] \nonumber \\
  &\qquad\quad\times\ddp{^2\Phia}{z^2}(\Rg,0)-\delta_{|j|1}\frac{\Rg}{2}e\ddp{^3\Phia}{R\partial z^2}(\Rg,0)\Bigg\}.
\end{align}
We can now rewrite $f_1$ (Eq.~\ref{eq:f1}) as
\begin{equation}\label{finalf1}
  f_1=f_{1,R}+f_{1,z},
\end{equation}
where
\begin{equation}
  f_{1,R}\equiv\Rep\Bigg\{\sum_{j=-1}^{1} c^R_{j m 0}
  {\rm F}_{j m 0}\eexp^{\img \paresq{j\theta_R+m\pare{\theta_\phi-\Omp t}}}\Bigg\},
\end{equation}
\begin{equation}
  f_{1,z}\equiv\Rep\Bigg\{\sum_{j,l=-1}^{1} c^z_{j m 2l}
  {\rm F}_{j m 2l}\eexp^{\img \paresq{j\theta_R+m\pare{\theta_\phi-\Omp t}+2l\theta_z}}\Bigg\},
\end{equation}
where the Fourier coefficients $c_{j k l}^R$ and $c_{j k l}^z$ are
given by \Eq{eq:cR}, \Eq{eq:cz}, and
\begin{equation}\label{eq:F}
  {\rm F}_{j k l}(J_R,J_\phi,J_z)\equiv \frac{j\dddp{f_0}{J_R}+
    k\dddp{f_0}{J_\phi}+l\dddp{f_0}{J_z}}{j\kappa+k\pare{\omega_\phi-\Omp}+l\nu}.
\end{equation}

\subsection{Moments of the distribution function}
One of the main motivation of the present work is to understand the
present response of a disc stellar population, represented by a DF
$f_0$ in an axisymmetric potential, to a quasi-static perturbing
non-axisymmetric potential in terms of radial and vertical mean
motions (e.g., \citealt{Faure2014}). Such mean motions can be computed
through the zeroth and first order moments of the perturbed DF
$f=f_0+\epsilon f_1$. Here we will assume a given form for $f_0$,
namely the Schwarzschild DF of \Eq{schwarzschild}.

We will focus on the mean motions projected on the plane (for the
radial motion) and on both sides of the plane (for the vertical
motion). Indeed, it was already shown numerically (\citealt{Faure2014,
  Monari2015}) that spiral or bar perturbations typically lead to a
{\it breathing mode} response of the disc, i.e. a density response
that has even parity with respect to the Galactic plane (i.e., is
plane-symmetric), and a mean vertical velocity field that has odd
parity. Hence we will concentrate hereafter on the projected surface
density $\Sigma(R,\phi)$, the projected mean radial velocity field
$\avvR(R,\phi)$, and the difference between the mean vertical velocity
field above and below the plane
\begin{equation}
\dvz(R,\phi) \equiv \avvz(z>0) - \avvz(z<0).
\end{equation}

These can be computed by integrating the perturbed DF over all $z$ (or
half of them in the case of the vertical motion) and all velocities,
i.e.,
\begin{subequations}\label{eq:mom1}
\begin{align}
  \Sigma(R,\phi)&\equiv\int_{-\infty}^{\infty}\de z\int\de^3\bv (f_0+\epsilon f_1), \\
  \Sigma(R,\phi)\avvR(R,\phi)&\equiv\int_{-\infty}^{\infty}\de z\int\de^3\bv v_R (f_0+\epsilon f_1), \\
  \Sigma(R,\phi)\dvz(R,\phi)&\equiv 4\int_0^{\infty}\de z\int\de^3\bv v_z (f_0+\epsilon f_1),
\end{align}
\end{subequations}
where
\begin{equation}
  \de^3\bv=\de v_\phi\de v_R\de v_z.
\end{equation}
Note that, by integrating over half of the $z$ for $\dvz$, we get only
half of the surface density, and have to multiply by two again to get
the subtraction between the mean vertical velocities above and below
the plane, hence the factor of four. Now, using the parity of the
functions, \Eq{eq:mom1} simplify to
\begin{subequations}\label{eq:mom2}
\begin{align}
  \Sigma(R,\phi)&=\int_{-\infty}^{\infty}\de z\int\de^3\bv~(f_0+\epsilon f_1), \\
  \Sigma(R,\phi)\avvR(R,\phi)&=\epsilon\int_{-\infty}^{\infty}\de z\int\de^3\bv~v_R~f_1, \\
  \Sigma(R,\phi)\dvz(R,\phi)&=4\epsilon\int_0^{\infty}\de z\int\de^3\bv~v_z~f_{1,z}.
\end{align}
\end{subequations}
These integrals have to be solved at constant $(R,\phi,t)$. To compute
the integrals over all velocities, we pass from the integration
coordinates $(v_R,v_\phi,v_z)$ (where $v_R$ and $v_z$ range from
$-\infty$ to $\infty$, and $v_\phi$ from $0$ to $\infty$) to
$(\theta_R,\theta_z,J_\phi)$ (where $\theta_R$ and $\theta_z$ range
from $-\upi/2$ to $\upi/2$, and $J_\phi$ from $0$ to $\infty$) via the
transformations
\begin{subequations}
\begin{align}
  v_R&=-\kappa(R-\Rg)\tan\theta_R, \\
  v_\phi&=J_\phi/R, \\
  v_z&=-\nu z\tan\theta_z,
\end{align}
\end{subequations}
and
\begin{subequations}
\begin{align}
  J_R&=\frac{(R-\Rg)^2\kappa}{2\cos^2\theta_R},\\
  J_z&=\frac{z^2\nu}{2\cos^2\theta_z},\\
  \theta_\phi&=\phi+\Delta\phi(\theta_R).
\end{align}
\end{subequations}
The Jacobian of the transformation is
\begin{equation}
  \de v_R \de v_\phi \de v_z = 
  \frac{\kappa\nu(R-\Rg)z}{R\cos^2\theta_R \cos^2\theta_z} 
  \de\theta_R \de\theta_z \de J_\phi,
\end{equation}
Using these transformations, as well as the approximations
$\omega_\phi\approx\Omega$, $\exp(\img m\Delta\phi)\approx(1+\img
m\Delta\phi)$, $\Delta\phi\approx -\gamma/\Rg
\sqrt{2J_R/\kappa}\sin\theta_R$ (i.e., up to the first order in $e$),
we compute the integrals of \Eq{eq:mom2}, and the DF $f_0$ of
\Eq{schwarzschild}, to obtain\footnote{Actually, the explicit results
  of \Eqs{eq:den}{eq:avz} are valid not only for the Schwarzschild DF
  of \Eq{schwarzschild} but for any DF that has a dependence on $J_R$
  and $J_z$ of the form ${\rm exp} \left(- \frac{J_R\kappa}{\sigmaR^2}
    - \frac{J_z\nu}{\sigmaz^2} \right)$.}
\begin{subequations}\label{eq:den}
\begin{align}
  \Sigma&=\Sigma_0+\epsilon\Sigma_1, \\
 {\rm where} \nonumber \\
  \Sigma_0&=\frac{(2\upi)^{3/2}}{R}\int_0^\infty\de J_\phi \frac{\sigmaR\sigmaz^2}{\nu}
  \eexp^{\frac{\psi_R}{\sigmaR^2}}f_0(0,J_\phi,0), \\
  \Sigma_1&=\Rep\Bigg\{\frac{(2\upi)^{3/2}\eexp^{\img\hphi}}{R}
  \sum_{j=-1}^1\int_0^\infty\de J_\phi
  \frac{\sigmaz}{\nu}\eexp^{\frac{\psi_R}{\sigmaR^2}}\pare{\hS_{j}+\hS_{j 0}'} \nonumber\\
      &\quad\times\paresq{\delta_{j0}-\delta_{|j|1}
      \pare{\delta_R-j\frac{m\gamma\sigmaR^2}{\kappa^2\Rg^2}}} \Bigg\}.
\end{align}
\end{subequations}
For the mean radial velocity, we get
\begin{align}\label{eq:avR}
  \Sigma&\avvR=\Rep\Bigg\{-\img\epsilon\frac{(2\upi)^{3/2}\eexp^{\img\hphi}}{R}
  \sum_{j=-1}^1\int_0^\infty\de J_\phi
  \frac{\sigmaz\sigmaR^2}{\nu\kappa\Rg}\eexp^{\frac{\psi_R}{\sigmaR^2}} \nonumber\\
      &\quad\times\pare{\hS_{j}+\hS_{j 0}'}\paresq{\delta_{j0}m\gamma-\delta_{|j|1}
      \pare{j+m\gamma\delta_R}}\Bigg\}.
\end{align}
Finally, for the difference of mean vertical velocities above and below the plane, we get
\begin{align}\label{eq:avz}
  \Sigma&\dvz=\Rep\Bigg\{-\img\epsilon\frac{8\upi\eexp^{\img\hphi}}{R}
  \sum_{j=-1}^1\sum_{l=-1}^1\int_0^\infty\de J_\phi
  \frac{\sigmaz^2}{\nu}\eexp^{\frac{\psi_R}{\sigmaR^2}}\frac{l}{2^{|l|}} \nonumber\\
      &\quad\times\hS_{j 2l}'\paresq{\delta_{j0}-\delta_{|j|1}
      \pare{\delta_R-jm\gamma\frac{\sigmaR^2}{\kappa^2\Rg^2}}}\Bigg\},
\end{align}
where
\begin{subequations}
\begin{align}
  \delta_R&\equiv\frac{R-\Rg}{\Rg}, \\
  \hphi&\equiv m\pare{\phi-\Omp t}, \label{hatphi} \\
  \psi_R&\equiv-\frac{\kappa^2\pare{R-\Rg}^2}{2}\\
  \hS_{j}&\equiv \sigmaR\sigmaz c^R_{jm0}\pare{\frac{\kappa\Rg^2}{2},J_\phi,0}{\rm F}_{jm0}\pare{0,J_\phi,0},\\
  \hS'_{jl}&\equiv \sigmaR\sigmaz c^z_{jml}\pare{\frac{\kappa\Rg^2}{2},J_\phi,\frac{\sigmaz^2}{\nu}}{\rm F}_{jml}\pare{0,J_\phi,0}.
\end{align}
\end{subequations}
The mean vertical stellar motion is thus non-zero because the factor
$\hS'_{j2l}$ is not the same for $l=1$ and $l=-1$ in the integrand of
\Eq{eq:avz}.

\section{Results}\label{sect:results}

\subsection{Spiral arm model}\label{sect:results_mod}

We now wish to obtain explicit results in the case of a given 3D
spiral arm perturbation of the Galactic potential. The parameters of
the axisymmetric Galactic potential $\Phi_0$, the equilibrium DF
$f_0$, and the spiral perturbation $\Phi_1$ can all be varied in order
to get different responses for different parameters, and they could
all be used as free parameters when fitting a distribution function to
observed stellar kinematics from large Galactic surveys. Here we fix
these parameters in order to illustrate the typical behavior of $f_1$.

For $\Phi_0(R,z)$, we choose a realistic potential for the Galaxy,
namely the Model~I of \citet{BT2008}, fitting several observed
properties of the Milky Way \citep[see also][]{DehnenBinney1998}. It has a spheroidal dark halo and bulge,
as well as three components for the disc potential: thin, thick, and
ISM disc. The disc densities decrease exponentially with both
Galactocentric radius and height from the Galactic plane.

For $f_0(J_R,J_\phi,J_z)$, we choose the Schwarzschild DF of
\Eq{schwarzschild} with $\Rd=2\Kpc$, $z_0=0.3\Kpc$, and
\begin{subequations}
\begin{align}
  \sigmaR(R)&=\sigmaR(R_0)\eexp^{-\frac{R-R_0}{5\Rd}},\\
  \sigmaz(R)&=\sigmaz(R_0)\eexp^{-\frac{R-R_0}{5\Rd}},
\end{align}
\end{subequations}
where $\sigmaR(R_0)=35\kmsec$, and $\sigmaz(R_0)=15\kmsec$.

As a perturbation $\Phi_1(R,z,\phi,t)$ we wish to use a tightly-wound
logarithmic spiral. Expressing an analytic potential-density pair for
a 3D spiral is not trivial. For instance, if we consider a logarithmic
spiral with radial wavenumber $k(R)=m/(R \, {\rm tan} \, p)$, where
$p$ is the pitch angle, one could multiply the 2D potential by ${\rm
  exp}(-|k z|)$, but this would have the drawback that the vertical
force field would be discontinuous in the plane. Instead, we use here
the spiral arms potential-density pair of \citet{CoxGomez2002}, which
closely resembles arms with a $\sech^2$ vertical fall-off. With this
potential-density pair, our $\Phia(R,z)$ in \Eq{eq:phi1} corresponds
to a logarithmic spiral with radially-varying amplitude and
radially-varying scale-height, which reads
\begin{equation}\label{eq:sppot}
  \epsilon \Phia(R,z)=-\frac{A}{\Rs K
    D}\eexp^{\img m\left[-\phis+\frac{\ln(R/\Rs)}{\tan
        p}\right]}\left[\sech\left(\frac{Kz}{\beta}\right)\right]^\beta,
\end{equation}
where
\begin{subequations}
\begin{align}
  K(R)&=\frac{2}{R\sin p}, \\
  \beta(R)&=K(R)\hs\paresq{1+0.4K(R)\hs}, \\
  D(R)&=\frac{1+K(R)\hs+0.3\paresq{K(R)\hs}^2}{1+0.3K(R)\hs}.
\end{align}
\end{subequations}
For the length and height parameters of this spiral potential, we
choose $\Rs=1\Kpc$ and $\hs=0.1\Kpc$. We also fix a phase
$\phis=-26\degr$ and consider our following results at present time
$t=0$. The spiral is chosen to be tightly-wound with $p=-9.9\degr$,
and the amplitude parameter is chosen to be $A=683.7\kmsect$. Finally,
we choose to consider a 2-armed spiral with $\Omp=18.9\kmseckpc$, so
that the main resonances are relatively away from the Solar
neighbourhood. The inner Lindblad resonance would be hidden in the
central bar region of the Galaxy (${\rm ILR}=1.56\Kpc$) and the
corotation is in the outer galaxy (${\rm CR}=11.49\Kpc$). These
parameters have been partly inspired by the 2D spiral potential
considered in \citet{Siebert2012}, and produce at $(R,z)=(8\Kpc,0)$ a
maximum radial force of the spiral that is $1\%$ of the force due to
the axisymmetric background.

With this form of the background potential, axisymmetric equilibrium
distribution function, and spiral potential, we can now compute the
Fourier coefficients of Eqs.~(\ref{eq:cR}) and (\ref{eq:cz}), as well
as the perturbed distribution function of \Eq{finalf1} and its moments
of \Eqs{eq:den}{eq:avz}.

\subsection{Moments of the distribution function}\label{sect:results_mom}
\subsubsection{Radial velocity gradient}

\begin{figure*}
  \centering \includegraphics[width=\columnwidth]{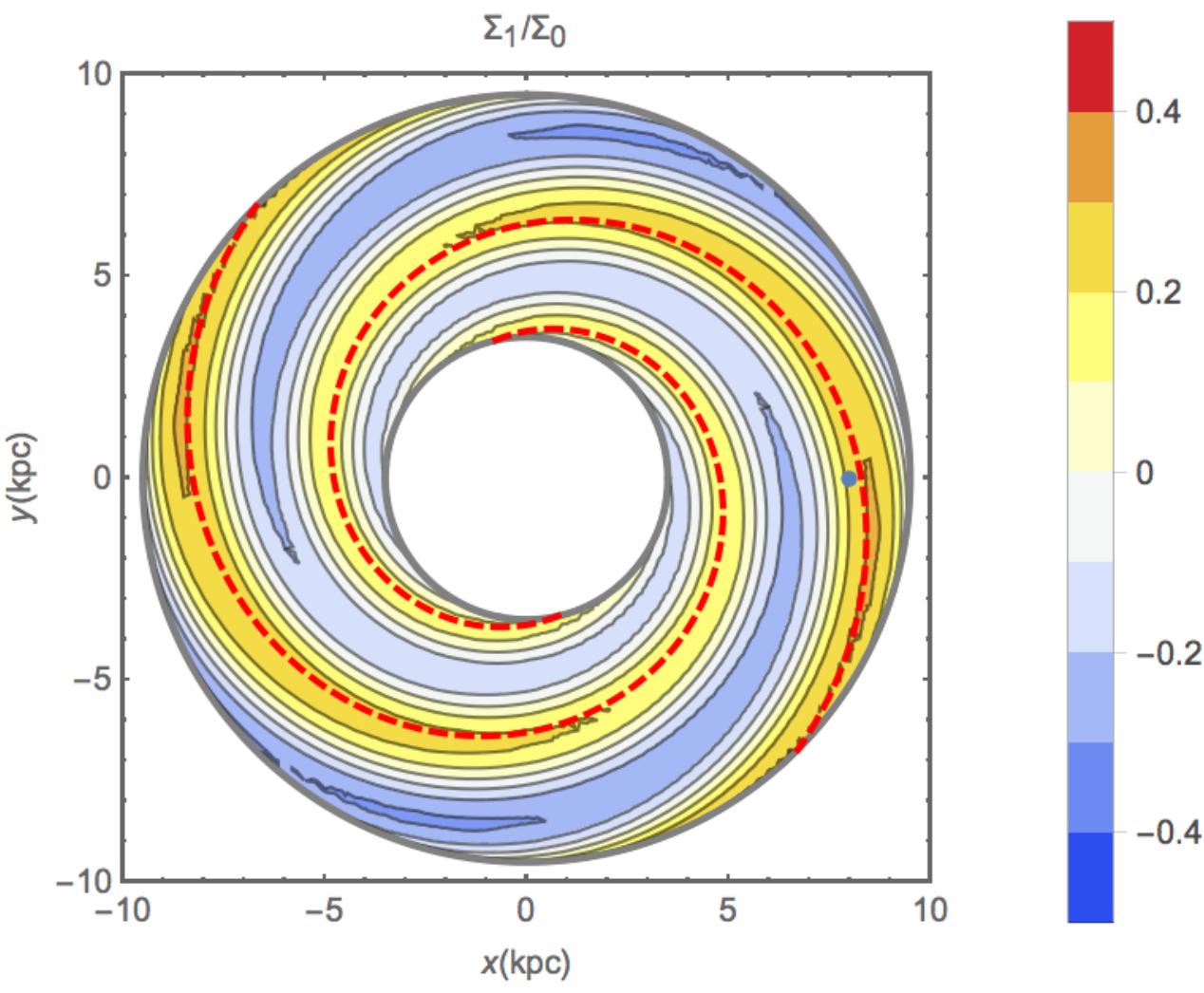}
  \includegraphics[width=\columnwidth]{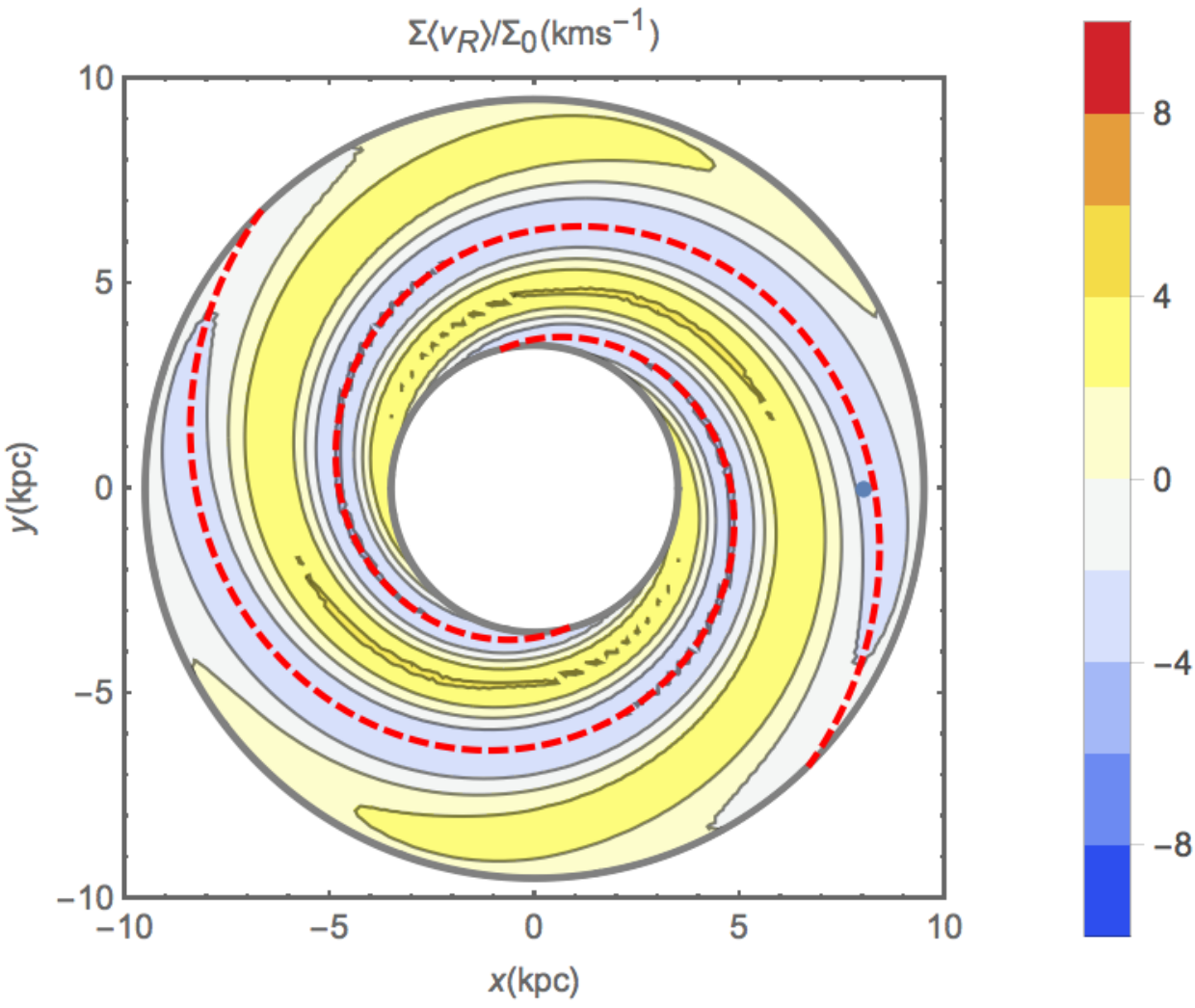}
  \includegraphics[width=\columnwidth]{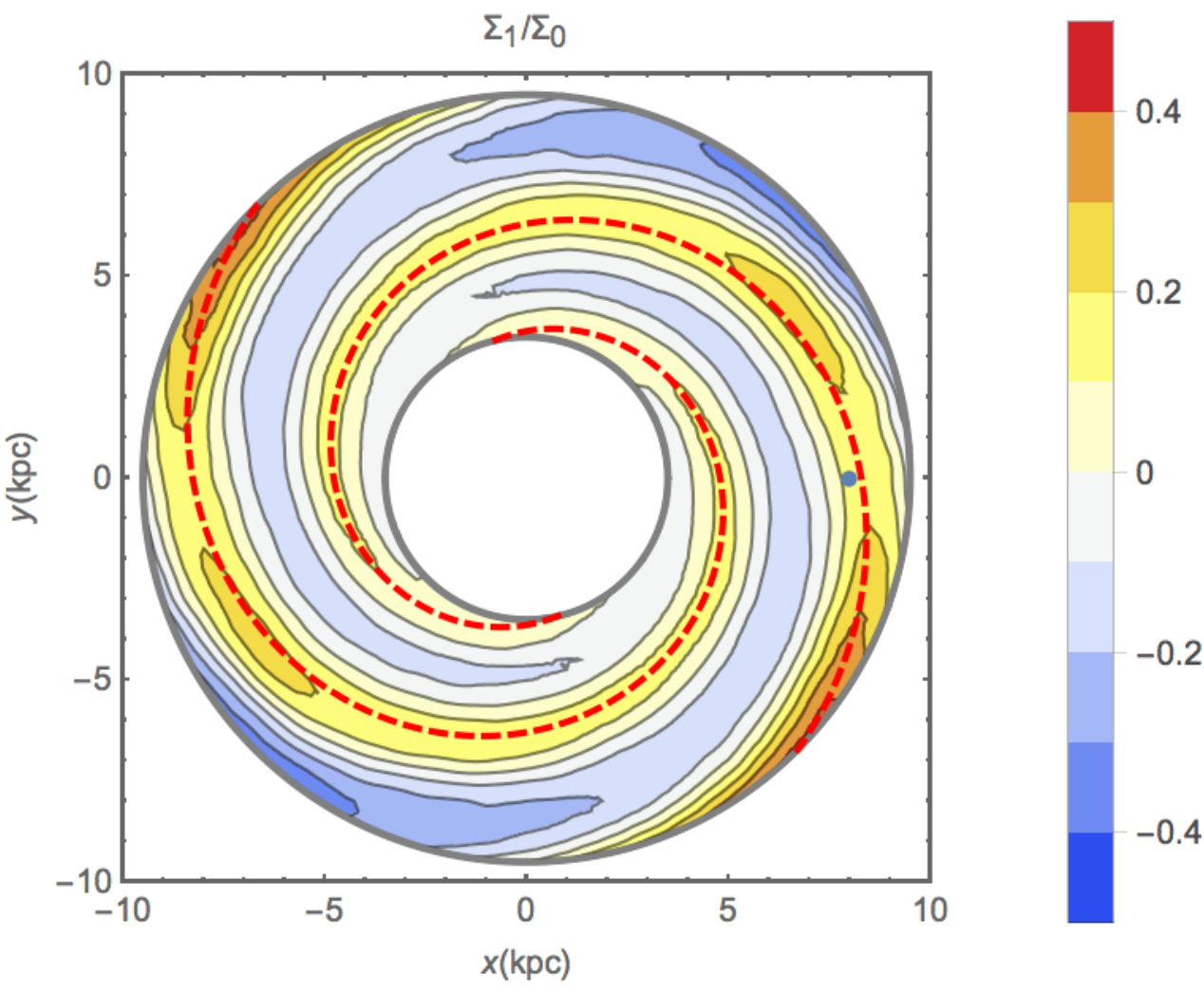}
  \includegraphics[width=\columnwidth]{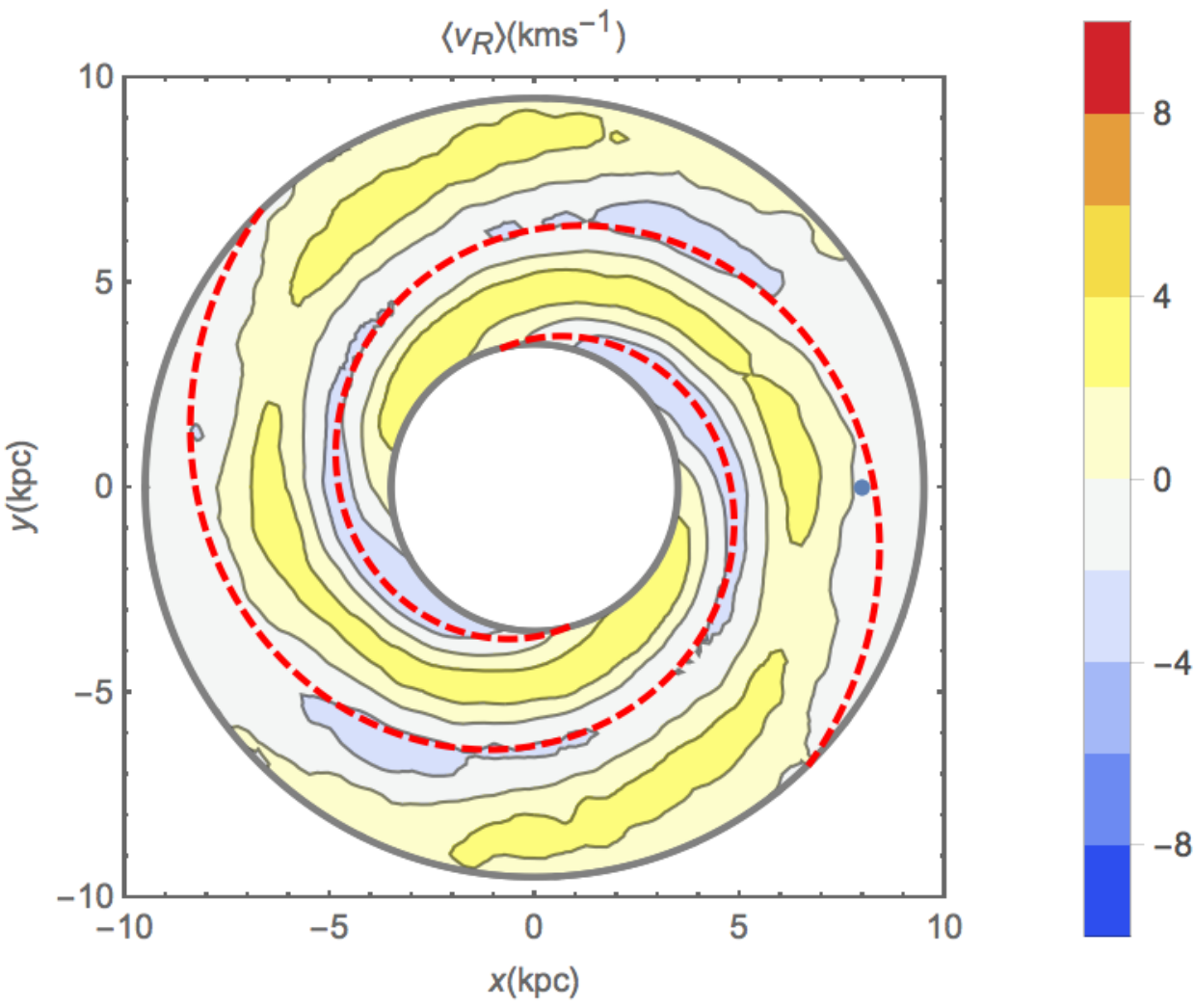}
  \caption{Moments induced by the potential perturbation \Eq{eq:sppot}
    on the \protect\cite{BT2008} Model~I potential. Top left:
    density wake $\Sigma_1/\Sigma_0$ obtained from \Eq{eq:den}. Top
    right: average radial speed $\Sigma\avvR/\Sigma_0$ obtained from
    \Eq{eq:avR}. Bottom left: density wake $\Sigma_1/\Sigma_0$
    obtained from the simulation. Bottom right: average radial speed
    $\avvR$ obtained from the simulation. $\Sigma_0$ is computed in
    the simulation averaging $\Sigma$ over $\phi$ at a certain
    $R$. The dashed red curves represent the loci of the arms.}
  \label{fig:thsim1}
\end{figure*}
\begin{figure*}
  \centering
  \includegraphics[width=\columnwidth]{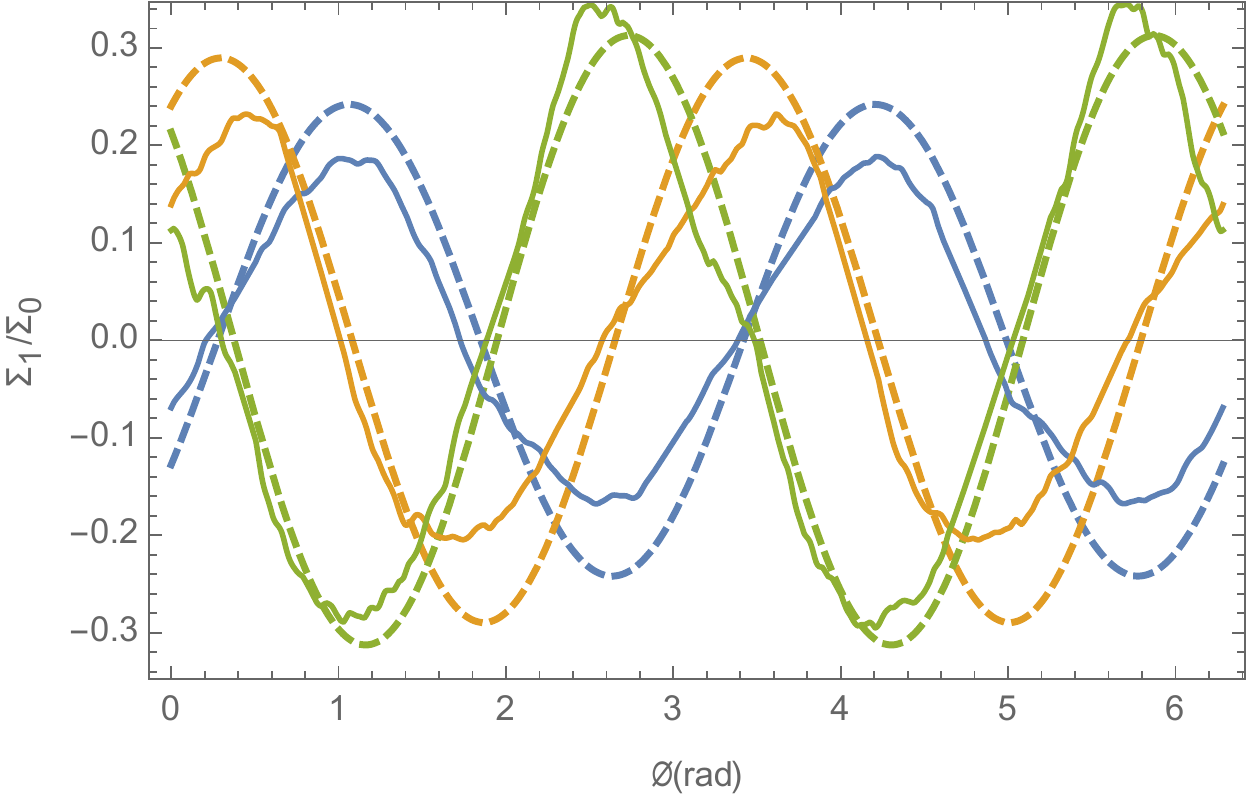}
  \includegraphics[width=\columnwidth]{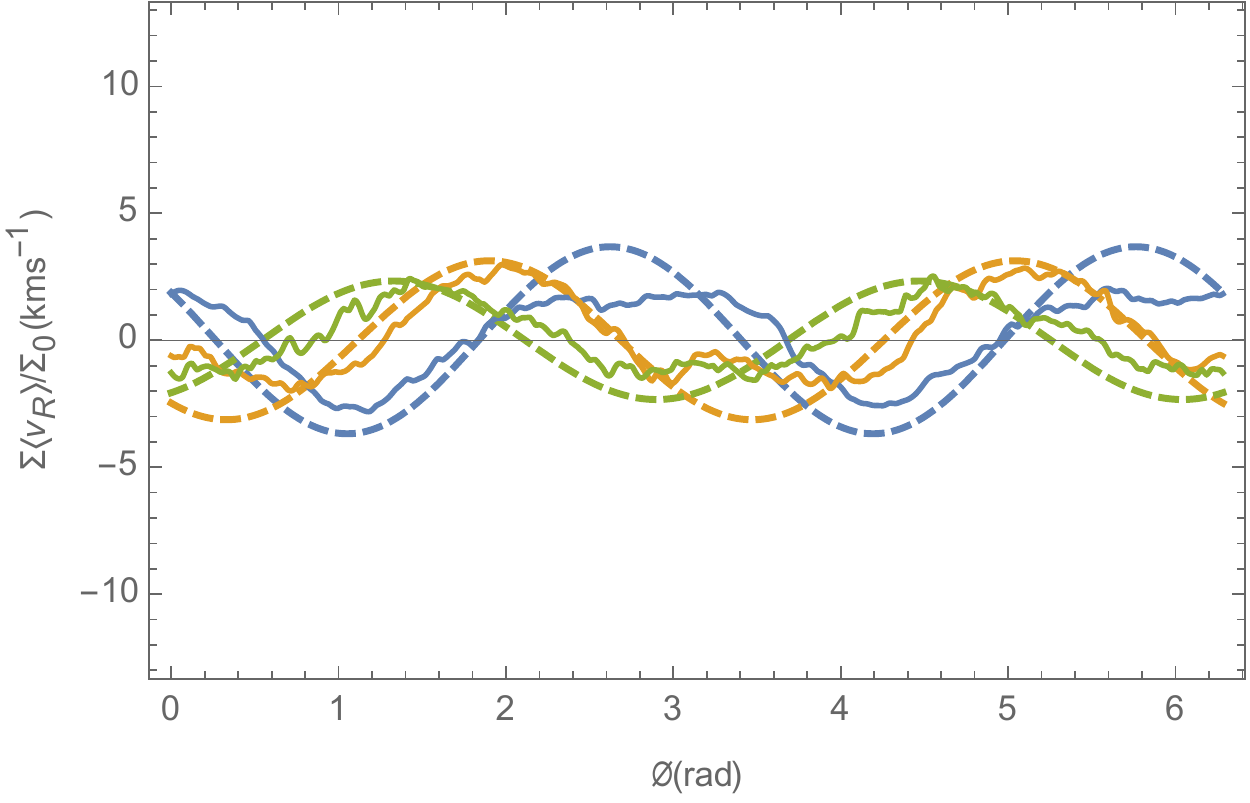}
  \caption{Comparison between the moments induced by the potential
    perturbation \Eq{eq:sppot} on the \protect\cite{BT2008} Model~I
    potential as a function of $\phi$ at three different radii
    computed with a numerical test-particle simulation (solid lines)
    and \Eqs{eq:den}{eq:avR} (dashed lines). Left panel:
    $\Sigma_1/\Sigma_0$. Right panel: $\avvR$. Blue lines:
    $R=7\Kpc$. Orange lines: $R=8\Kpc$. Green lines: $R=9\Kpc$. The
    moments of the simulation are computed inside $x-y$ square bins of
    $0.25\Kpc$ side, smoothed with a Gaussian filter on a scale of
    $0.5\Kpc$, and polynomial interpolated on the $x-y$
    grid.}\label{fig:comp1}
\end{figure*}
\begin{figure}
  \centering
  \includegraphics[width=\columnwidth]{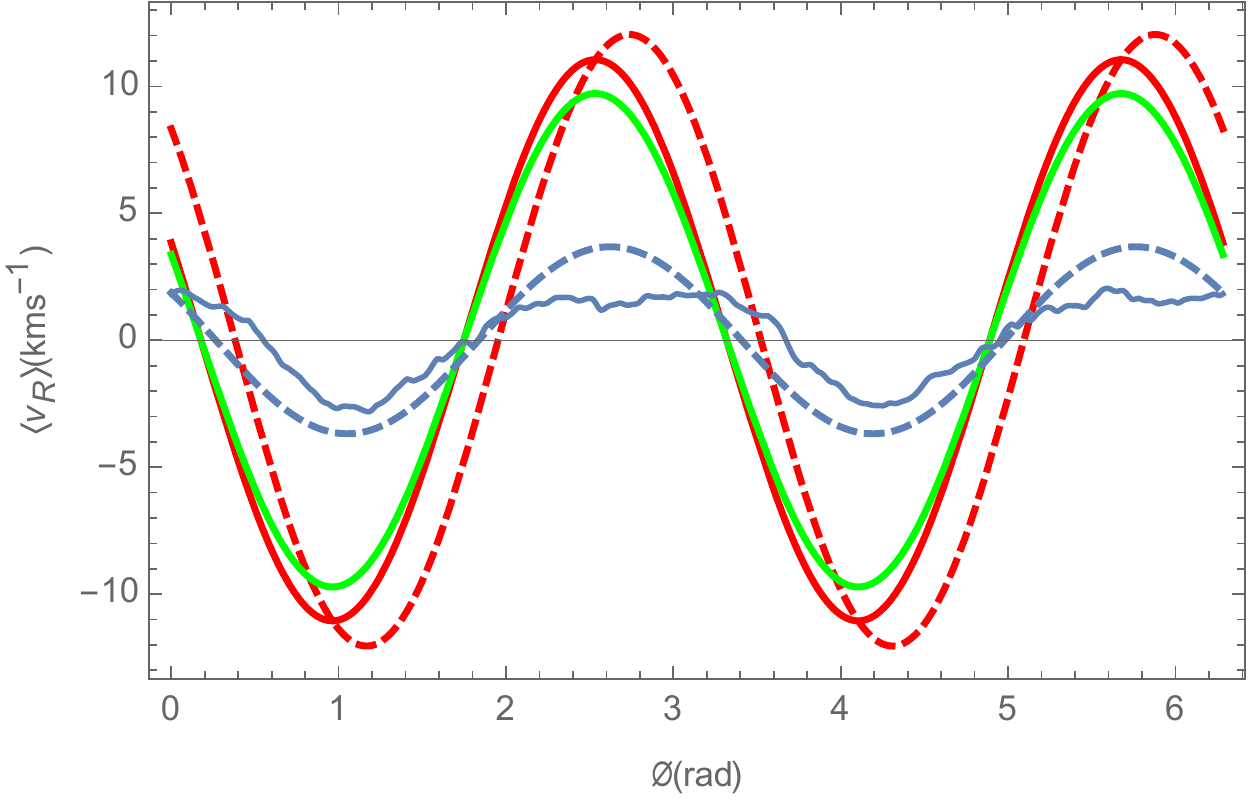}
  \includegraphics[width=\columnwidth]{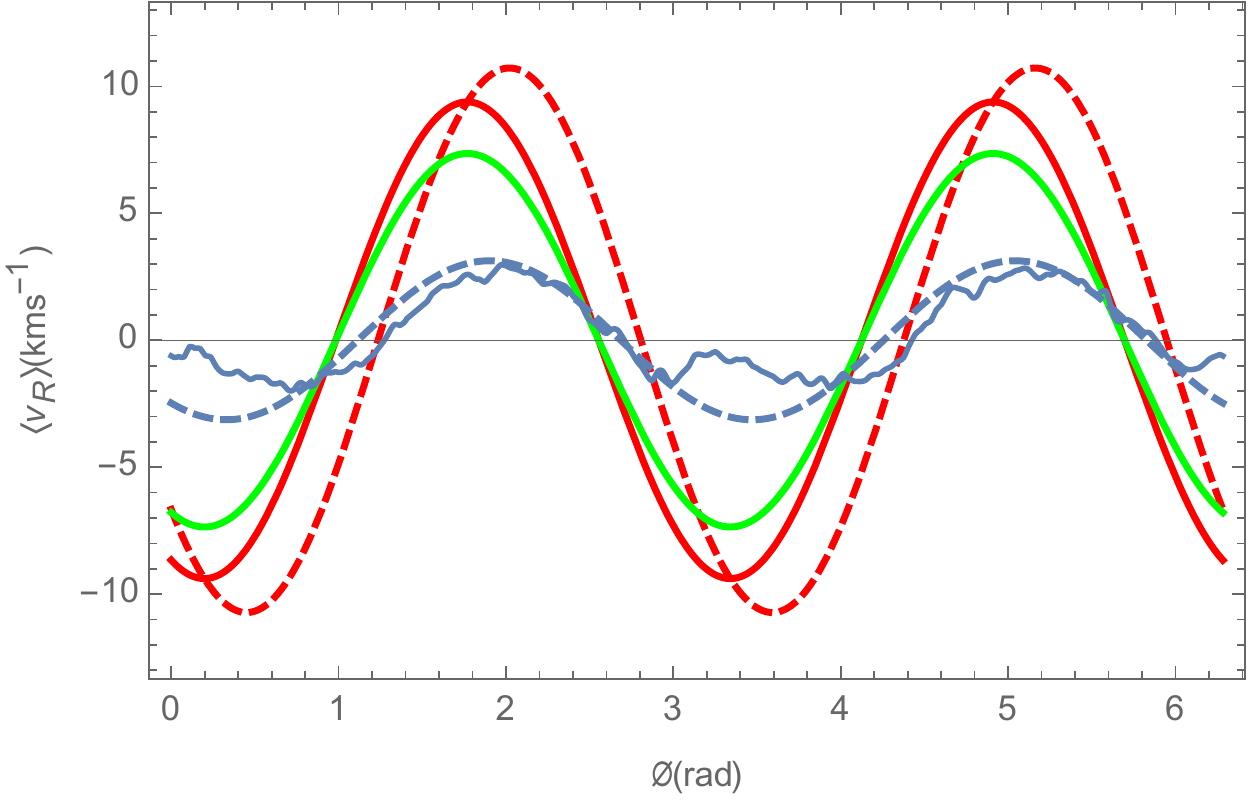}
  \includegraphics[width=\columnwidth]{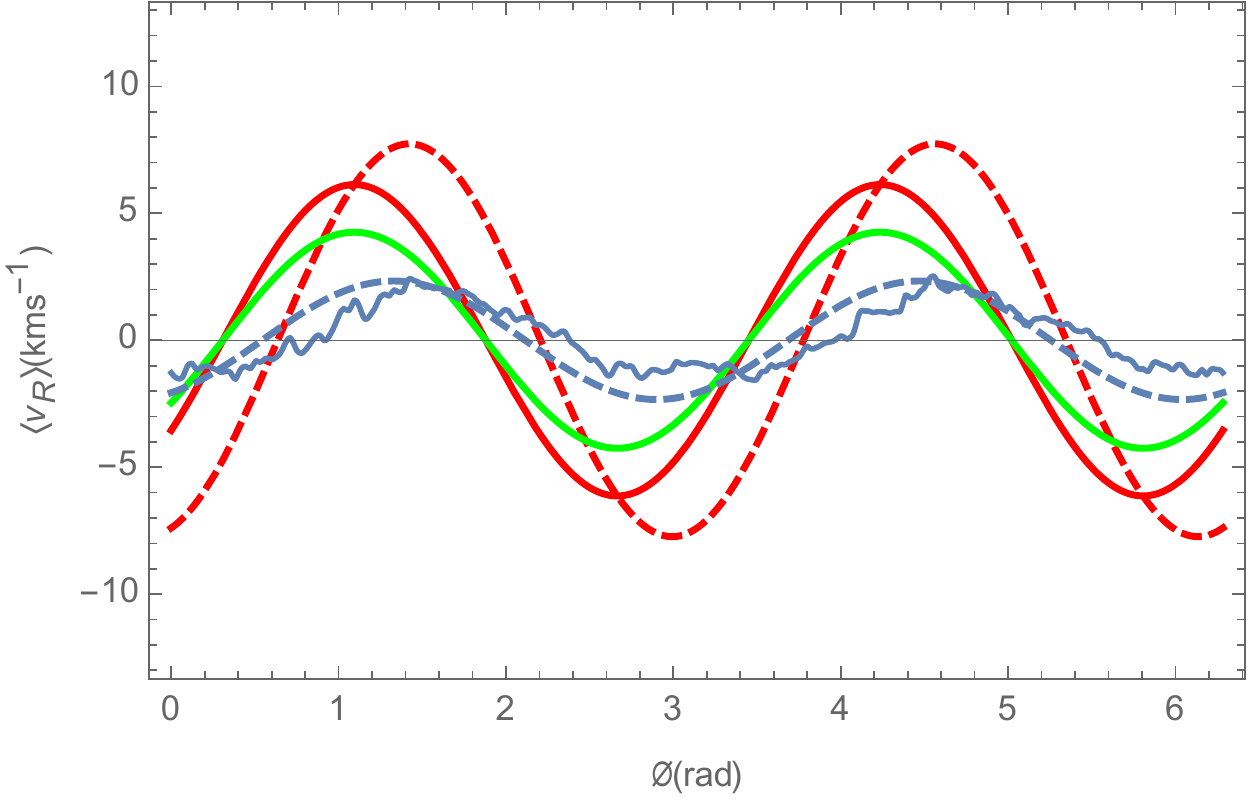}
  \caption{Several predictions for the response of mean $v_R$ of a
    stellar system or a cold fluid to the potential model used in this
    work. Predictions by \protect\citep[predictions
      by][]{LinShu,LinShu1966,LinShu1969}: red dashed line
    \Eq{eq:uRa}, red solid line \Eq{eq:uRas}, green line \Eq{eq:uRas}
    multiplied by the reduction factor $\calF$ by
    \protect\citealt{BT2008}. Blue dashed line \Eq{eq:avR}, blue solid
    line simulation. Top panel: $R=7\Kpc$. Central panel:
    $R=8\Kpc$. Bottom panel: $R=9\Kpc$.  }\label{fig:comp2}
\end{figure}
\begin{figure*}
  \includegraphics[width=\columnwidth]{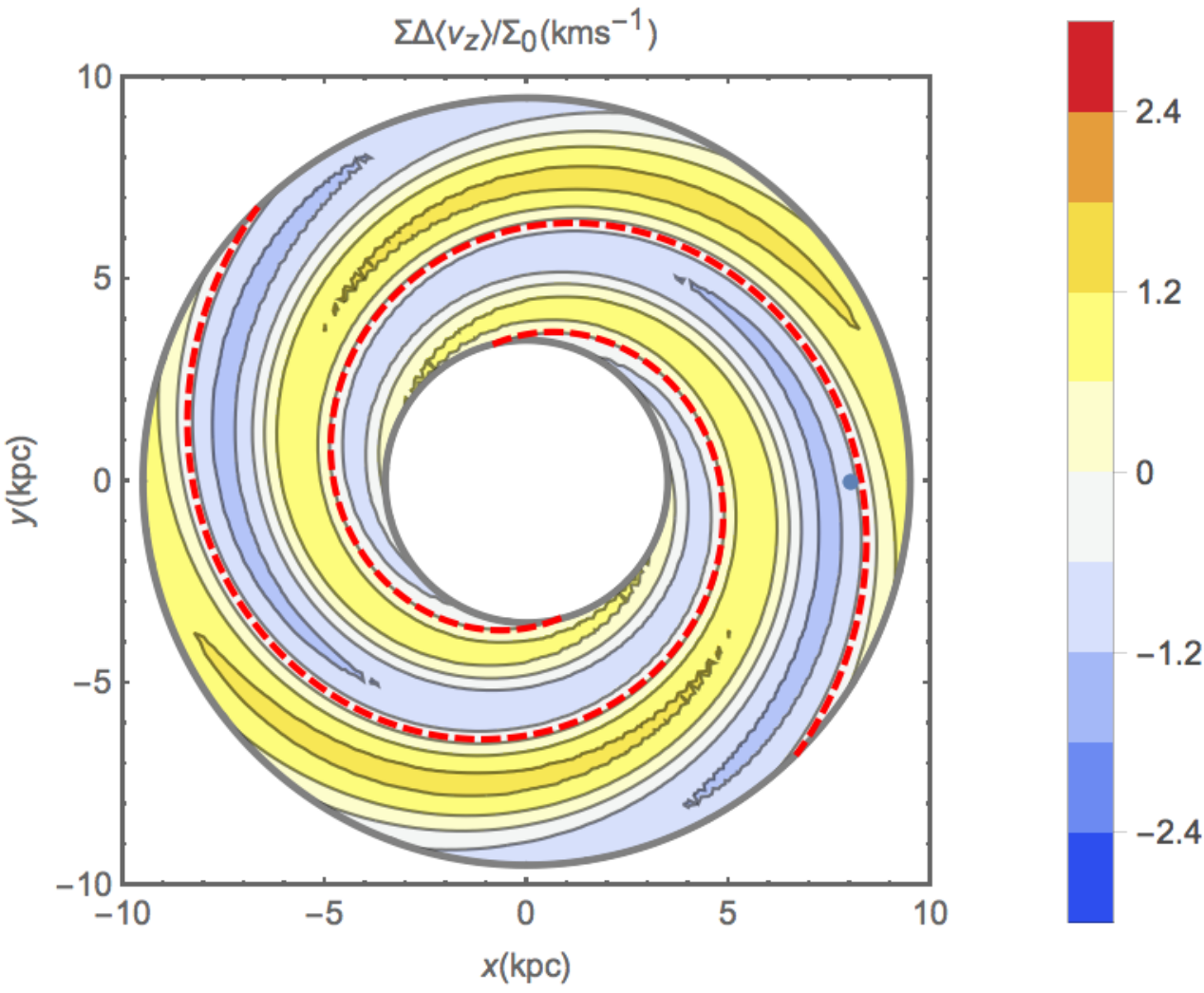}
  \includegraphics[width=\columnwidth]{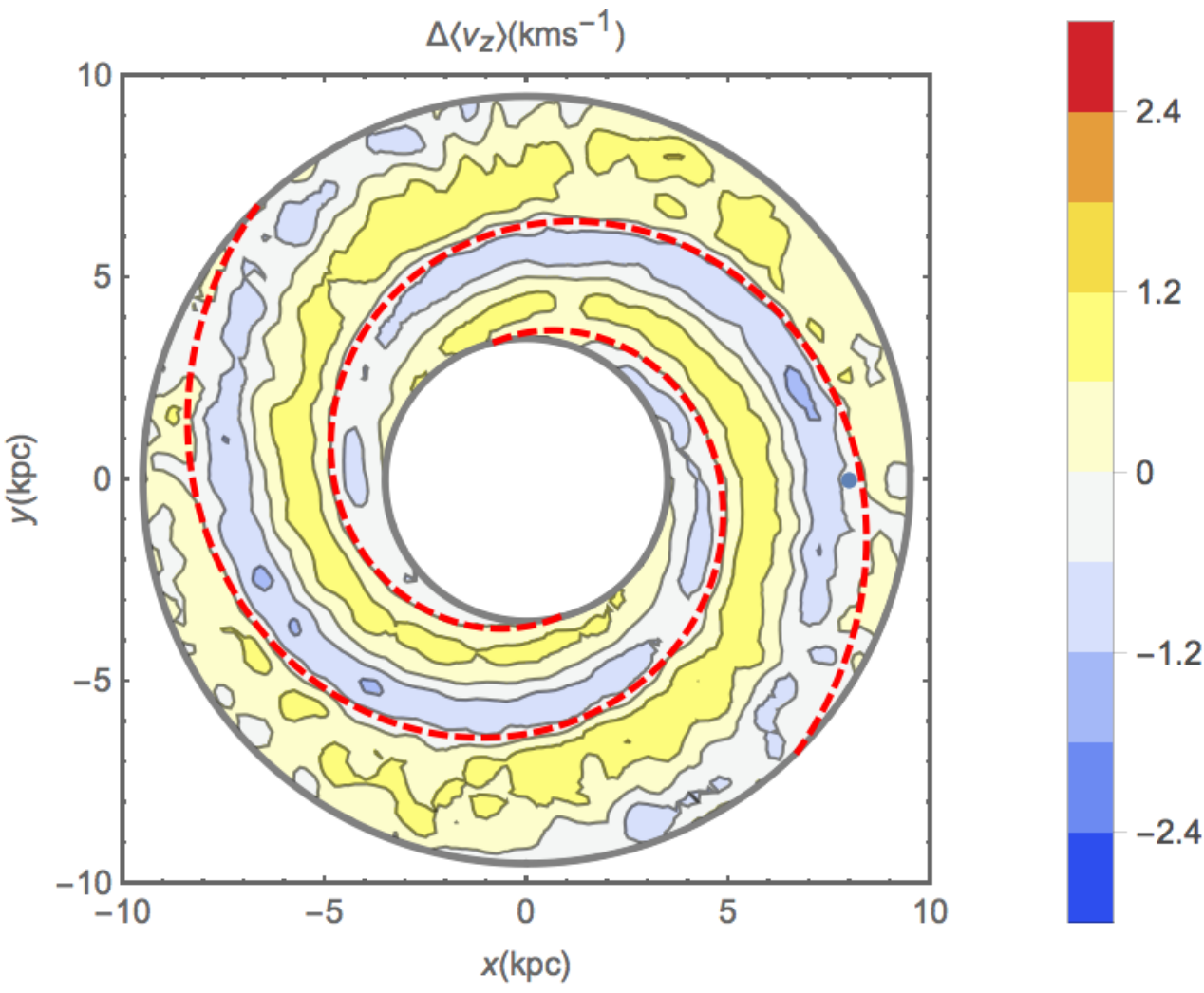}
  \caption{Mean $v_z$ motions induced by the potential perturbation
    \Eq{eq:sppot} on the \protect\cite{BT2008} Model~I
    potential. Left panel: north-south difference between the average
    vertical speed $\Sigma\dvz/\Sigma_0$ obtained from
    \Eq{eq:avz}. Right panel: north-south difference between the average
    vertical speed computed from the simulation. For the
    simulation $\dvz$ is computed inside $x-y$ square bins of
    $0.25\Kpc$ side, and smoothed with a Gaussian filter on a scale of
    $0.5\Kpc$ The dashed red curves represent the loci of the arms.}
  \label{fig:thsim2}
\end{figure*}
\begin{figure}
  \centering
  \includegraphics[width=\columnwidth]{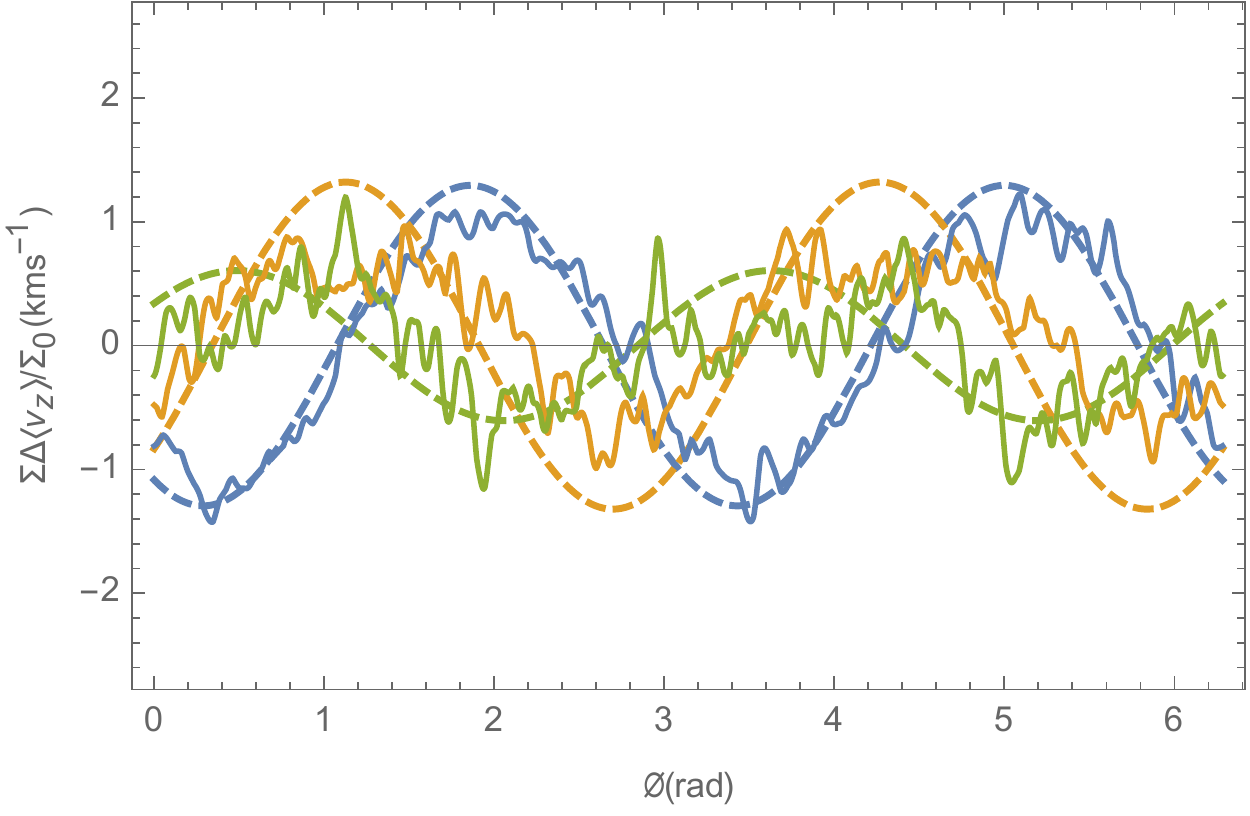}
  \caption{Comparison between the north-south difference in mean $v_z$
    motion $\dvz$ induced by the potential perturbation \Eq{eq:sppot}
    on the \protect\cite{BT2008} Model~I potential as a function of
    $\phi$ at three different radii computed with a numerical
    test-particle simulation (solid lines) and \Eq{eq:avz} (dashed
    lines). Blue lines: $R=7\Kpc$. Orange lines: $R=8\Kpc$. Green
    lines: $R=9\Kpc$. The quantity for the simulation is computed
    inside $x-y$ square bins of $0.25\Kpc$ side, smoothed with a
    Gaussian filter on a scale of $0.5\Kpc$, and polynomial
    interpolated on the $x-y$ grid.}\label{fig:comp3}
\end{figure}
\begin{figure}
  \centering
  \includegraphics[width=\columnwidth]{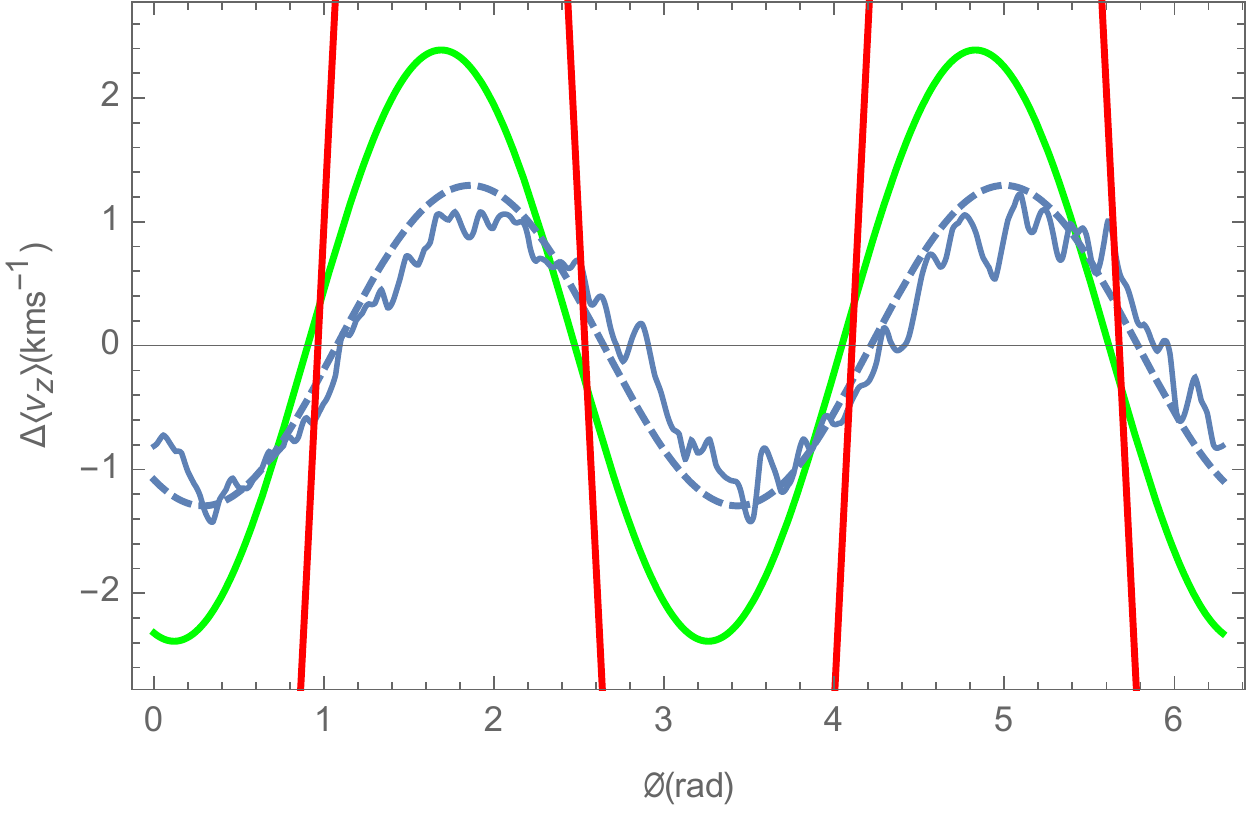}
  \includegraphics[width=\columnwidth]{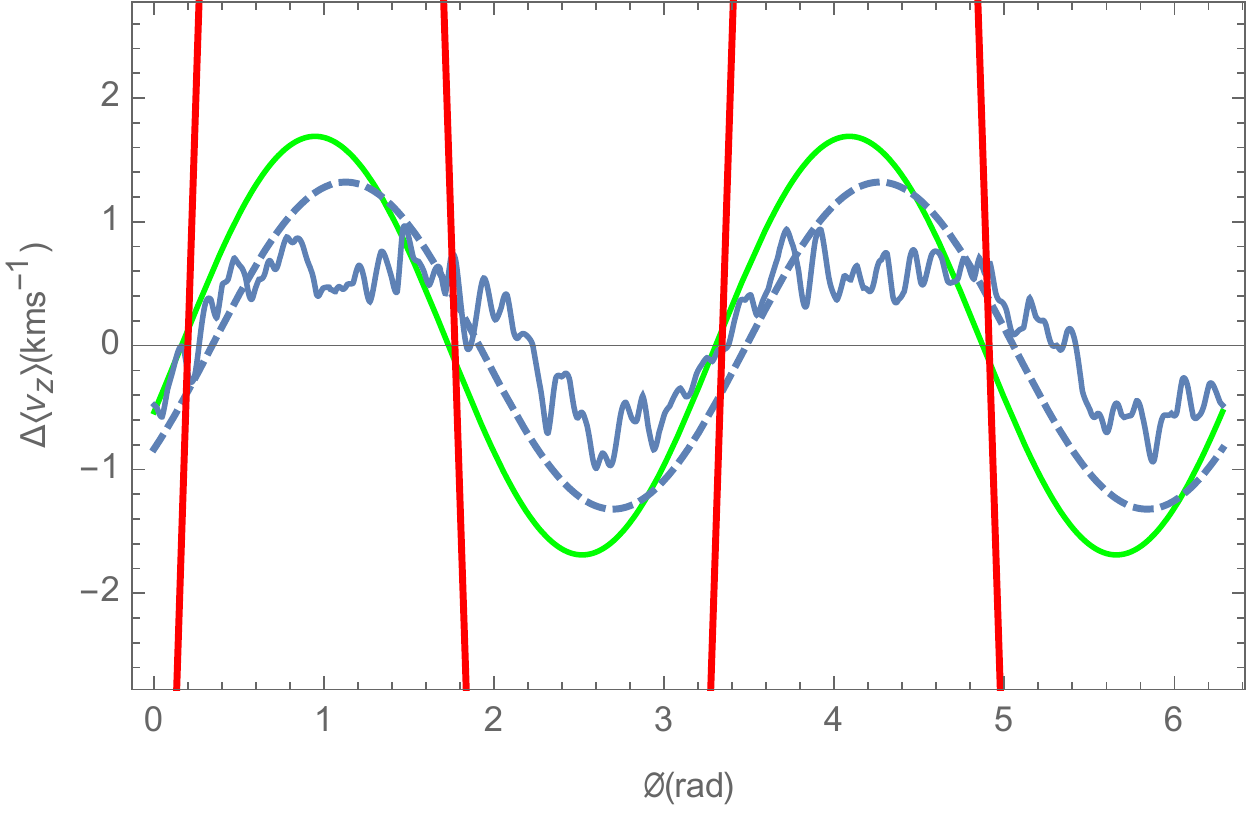}
  \includegraphics[width=\columnwidth]{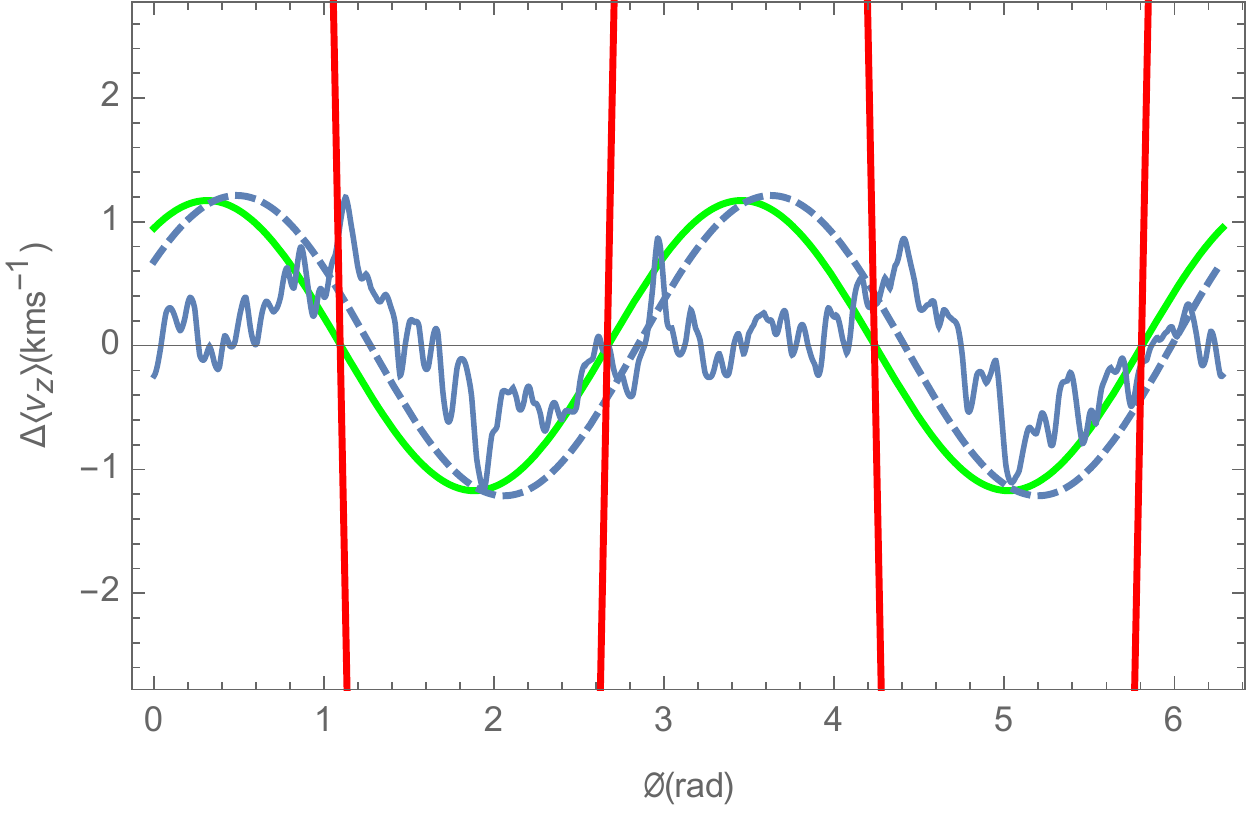}
  \caption{Several predictions for the response of mean $v_z$ of a
    stellar system or a cold fluid to the potential model used in this
    work (red solid line \Eq{eq:uza}, green line method by
    \citet{Monari2015} for a cold stellar disc, blue dashed line
    \Eq{eq:avz}, blue solid line simulation). Top panel:
    $R=7\Kpc$. Central panel: $R=8\Kpc$. Bottom panel:
    $R=9\Kpc$.}\label{fig:comp4}
\end{figure}

We first consider the integrals in \Eqs{eq:den}{eq:avR}, which have to
be computed numerically. In practice, for a given $R$, the integral on
$J_\phi$ is computed in the interval of angular momenta corresponding
to circular orbits at the radii where the circular velocity is $v_c
\pm 2 \sigmaR$ (we tested that the results obtained in this way are
stable on larger integration ranges). The moments are actually Fourier
modes themselves, i.e., they have the form
$q(R,\phi)=\Rep\{q_{\mathrm{a}}(R)\exp(\img\hphi)\}$ where $\hphi$ is
defined as in \Eq{hatphi}. We evaluate $q_{\mathrm{a}}(R)$ numerically
on a grid of $R$ values between $1$ and $10\Kpc$ with a step
$0.25\Kpc$, and use a 3rd order polynomial interpolation on this grid
to obtain the value $q_{\mathrm{a}}(R)$ at a generic $R$ point.

In \Fig{fig:thsim1} we plot $\Sigma_1/\Sigma_0$ and
$\Sigma\avvR/\Sigma_0$ as obtained from \Eqs{eq:den}{eq:avz}. As we
see, the maxima of the response density wake $\Sigma_1/\Sigma_0$
closely follow the loci of the spiral arm potential (dashed red
curves), as one expects. On the other hand, stars on the arms tend to
move towards the center of the Galaxy ($\avvR<0$), while those in the
interarm regions tend to move outside ($\avvR>0$).

In order to illustrate how our analytic calculations allow to
physically interpret the outcome of simulations, we compare the
moments induced by the perturbation derived analytically with those
computed with a numerical test-particle simulation. The initial
conditions are drawn from $f_0$, and with the same potential
$\Phi_0+\epsilon\Phi_1$ (where $\Phi_1$ grows slowly with time, until
it reaches the final amplitude used for the analytical
predictions). The details of this simulation can be found in
\cite{Monari2015}, where the only difference with the present
simulations is that, in that previous work, $\Phi_1$ was a bar
potential instead of the spiral arms that we use here. The results of
this simulation are depicted in \Fig{fig:thsim1}. We find a very good
agreement between the position of the maxima and minima of the moments
and the loci of the spiral arms. Moreover, the amplitude of the
perturbed density and motions appear to be similar to the analytical
predictions. A closer look to this comparison with the simulation is
presented in \Fig{fig:comp1}. Here the comparison is made at three
different radii, in the neighborhood of $R_0=8$~kpc: $R=7\Kpc$,
$R=8\Kpc$, and $R=9\Kpc$. These plots confirm the agreement between
the simulation and the analytical predictions. Some small
discrepancies are of course present, and are due to a combination of
different effects. One of them is the discrete nature of the
simulations, and the fact that they never reach complete
phase-mixing. The second is that, although the area covered is away
from the ILR and CR, there still are non-linear effects due to the
resonances of higher order than the Lindblad resonances that our
analytical method does not describe (e.g., due to the $4:1$ inner
ultra-harmonic resonance between $\Omega-\Omp$ and $\kappa$, which in
our case falls at $R=7.61\Kpc$). The third is the presence of very
eccentric orbits, especially in the inner regions of the Galaxy, while
\Eqs{eq:den}{eq:avR} are valid only for moderate eccentricities.

All this is especially interesting in view of the large-scale radial
velocity gradient first observed in the Galaxy by \citet{Siebert2011}
with the RAVE survey. This was interpreted as the possible effect of
either a $m=2$ spiral \citep{Siebert2012} or the Galactic bar
\citep{Monari2014}. In this respect, it is interesting to note that
the amplitude of the radial velocity fluctuations generated by our
spiral potential here are of the same order of magnitude as those
observed. It should however be noted that subsequently observed
large-scale line-of-sight velocity fluctuations with a few red clump
stars from the APOGEE survey seem to be more compatible with the
effect of the bar \cite{Bovy2015,Grand2015}.

In \citet{Siebert2012}, a comparison between the RAVE data and various
spiral models was made by using the traditional {\it reduction factor}
$\calF$ of \citet{LinShu,LinShu1966,LinShu1969} -- see also
\citet{BT2008}. In the case of a cold, pressureless fluid it can
indeed be shown that the linear response to a non-axisymmetric
rotating density perturbation $\epsilon\Phi_1$ in the radial velocity
on the Galactic plane is
\begin{equation}
  \epsilon \uRo (R,\phi)=\Rep\parec{\uRa (R) \eexp^{\img m(\phi-\Omp t)}},
\end{equation}
where
\begin{equation}\label{eq:uRa}
  \uRa(R)=\img\frac{m}{\Delta(R)} 
  \Bigg\{ \paresq{\Omp-\Omega(R)}\frac{\de \Phia}{\de R}(R,0)
   -\frac{2\Omega(R)\Phia(R,0)}{R}\Bigg\},
\end{equation}
and $\Delta(R)\equiv\kappa(R)^2-m^2\paresq{\Omp-\Omega(R)}^2$. When
the perturbing potential is a tightly wound spiral, the second term in
the r.h.s. of \Eq{eq:uRa} is much smaller than the first term, and can
be omitted, so that \Eq{eq:uRa} simplifies to
\begin{equation}\label{eq:uRas}
  \uRa(R)\approx\img\frac{m\paresq{\Omp-\Omega(R)}}{\Delta(R)}\frac{\de
    \Phia}{\de R}(R,0).
\end{equation}
\citet{LinShu,LinShu1966,LinShu1969} offer a way to rewrite
\Eq{eq:uRas} in the case of a stellar disc, i.e., by multiplying it by
a reduction factor $\calF$ whose derivation is reported in Appendix K
of \citet{BT2008}. In \Fig{fig:comp2} we compare all these predictions
with the \Eq{eq:avR} of this work and the outcome of our numerical
simulations at $R=R_0$. Since $\calF$ was derived for tightly wound
spirals only, we use the best fit tightly wound spiral potential with
the same pitch angle $p$ to $\epsilon\Phi_1$ of this work in the range
of $R$, $6\Kpc<R<8\Kpc$ (left panel), $7\Kpc<R<9\Kpc$ (cental panel),
and $8\Kpc<R<10\Kpc$ (right panel).  We notice that there is a
noticeable difference in the amplitude predicted by the Lin-Shu
approximation, even with the reduction factor, (a factor $\sim 2$ or
more), and the results obtained using \Eq{eq:avR} of the present work:
the latter case actually describes much better the numerical
simulation, calling for a re-investigation of non-axisymmetric
kinematic features in future surveys with our present DF-based method
rather than a simple reduction factor.  There are several likely
reasons for this difference. First of all, our approach is
three-dimensional, and takes explicitly into account the vertical
velocity dispersion of stars in the response to the
perturbation. Second, we do not neglect the tangential force term
which is usually neglected for tightly-wound spirals. Third, we use
the guiding radius to evaluate our quantities instead of the present
position which is used as a proxy in the Lin-Shu approach. Finally,
the Lin-Shu approach assumes for the time-variation of the azimuthal
angle that of a circular orbit, which is a good approximation only for
very small eccentricities. It is a combination of these effects which
leads to the present difference with the Lin-Shu reduction factor.

\subsubsection{Vertical bulk motions: breathing mode of the disc}
One of the immense advantages of working with a 3D spiral model is
that it allows us to investigate the effect of the spiral on mean
stellar {\it vertical} motions. This is especially interesting given
that recent Milky Way large spectroscopic surveys have consistently
indicated that the mean vertical motion of stars above and below the
plane was typically non-zero
(\citealt{Widrow2012,Williams2013,Carlin2013}). Such a behaviour was
originally associated uniquely with external excitations of the disc
by a passing satellite galaxy or a dark matter substructure
(\citealt{Widrow2012,Gomez2013,YannyGardner2013,Feldmann}). It is
however useful to separate such stellar bulk motions into two types of
vertical oscillations. If the density perturbation has odd parity with
respect to the Galactic plane, and the vertical velocity field has
even parity, the disc itself is subject to a corrugation pattern which
is called a ``bending mode". These are indeed mostly caused by
external perturbers \citep{Xu2015,Delavega2015,Gomez2015}. On the
other hand, if the density wake has even parity while the vertical
velocity field has odd parity (i.e., a rarefaction-compression
pattern), the oscillation is called a ``breathing mode". Such
breathing modes have been shown through test-particle simulations and
approximate analytical considerations to be natural consequences of
internal non-axisymmetries such as the bar and spiral arms
\citep{Faure2014,Monari2015}. The same effect was also found in
self-consistent simulations of isolated galaxies developing spiral
instabilities \citep{Debattista2014}. It was even shown that the
breathing mode present in the simulation of a Milky Way like galaxy
bombarded by satellites, which was analyzed by \citet{Widrow2014}, was
actually most probably linked to the bar formation rather than induced
by the satellites themselves \citep{Monari2015}.

Our present analytic calculations allow for the first time a rigorous
and fully dynamical understanding of spiral-induced breathing modes
away from the main resonances (and in the absence of resonance
overlaps of multiple patterns, which will be the topic of further
work). For this, it suffices to integrate \Eq{eq:avz} in a similar
manner as \Eq{eq:avR}. The resulting $\Sigma\dvz/\Sigma_0$ is plotted
on \Fig{fig:thsim2}. As can be seen, stars tend to vertically move
away from the Galactic plane at the outer edge of spiral arms
($\dvz>0$) and towards the plane at the inner edge ($\dvz<0$), with a
clear phase-shift w.r.t. the mean radial motion, already noted in
\citet{Faure2014}. Again, we compare this to the results of our
test-particle simulation (\Fig{fig:thsim2} and \Fig{fig:comp3}) and
find a good agreement.

If we assume again a pressureless fluid as in the 2D case, if we
additionally assume that $\lddp{\Phi_0}{z}\ll\lddp{\epsilon\Phi_1}{z}$
(which is of course a wrong assumption to make in the present case),
and that
\begin{equation}
  \epsilon \uzo (R,\phi,z)=\Rep\parec{\uza (R,z) \eexp^{\img m(\phi-\Omp t)}},
\end{equation}
the third of Euler's equation (without any reduction factor) to the
first order in $\epsilon$ leads to \citep[as first shown
  in][]{Faure2014}
\begin{equation}\label{eq:uza}
  \uza(R,z)=\img\frac{\lddp{\Phia(R,z)}{z}}{ m \paresq{\Omega(R)-\Omp}}.
\end{equation}
We compare \Eq{eq:uza} with the predictions of \Eq{eq:avz} in
\Fig{fig:comp4}. This comparison is made by averaging $u_z(R,\phi,z)$
over $z$ with weight $\exp\pare{-\nu^2z^2/2/\sigmaz^2}$ (i.e., in the
case where the vertical density is isothermal like in the case of
Schwarzschild's DF). The predictions of \Eq{eq:uza} are an order of
magnitude larger than the predictions of \Eq{eq:avz} and the
simulation (so much that we do not show, for readability, the complete
range of \Fig{fig:comp4}). The phases are instead in prefect
agreement. A more sophisticated (albeit not fully dynamical) approach
was taken by \cite{Monari2015}, relating the radial and tangential
motions for a very cold stellar disc or fluid to the vertical motions
via the continuity equation. The predictions of \cite{Monari2015}
(again, averaged along $z$ with weight the isothermal density
distribution) are also shown on \Fig{fig:comp4}, allowing to show the
typical reduction factor (as well as some phase-shift related to
missing terms in the cold fluid approximation). We note that the
breathing modes are qualitatively similar to those observed in the
extended solar neighbourhood \citep{Williams2013}, but that the
amplitude of these motions is much lower than observed. It
nevertheless remains to be seen how the coupling of multiple
perturbers will affect these vertical motions (Monari et al. in
prep.).

\subsection{Distribution function at a point in configuration
  space}\label{sect:results_DF}
\begin{figure}
  \centering
  \includegraphics[width=\columnwidth]{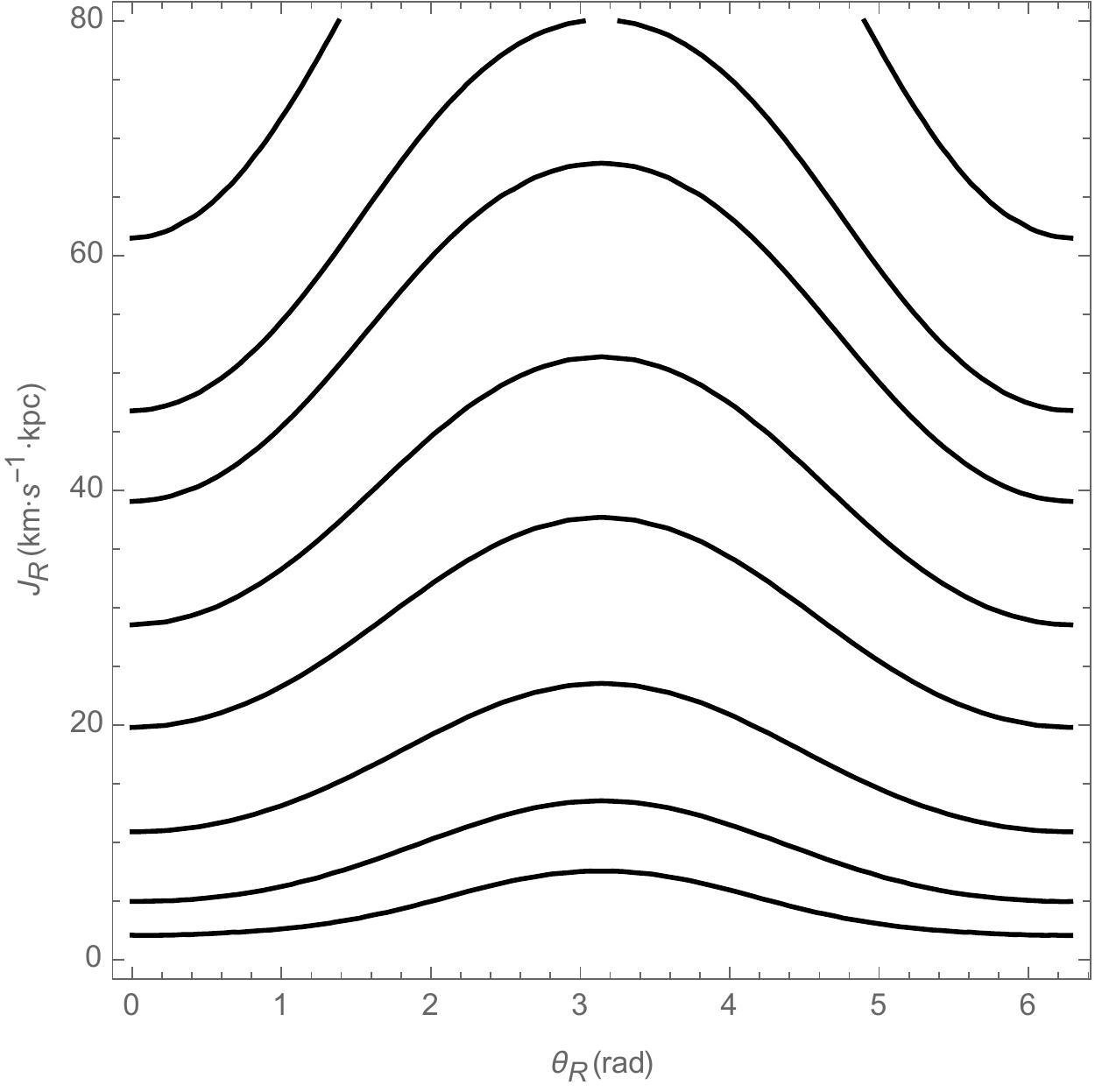}
  \includegraphics[width=\columnwidth]{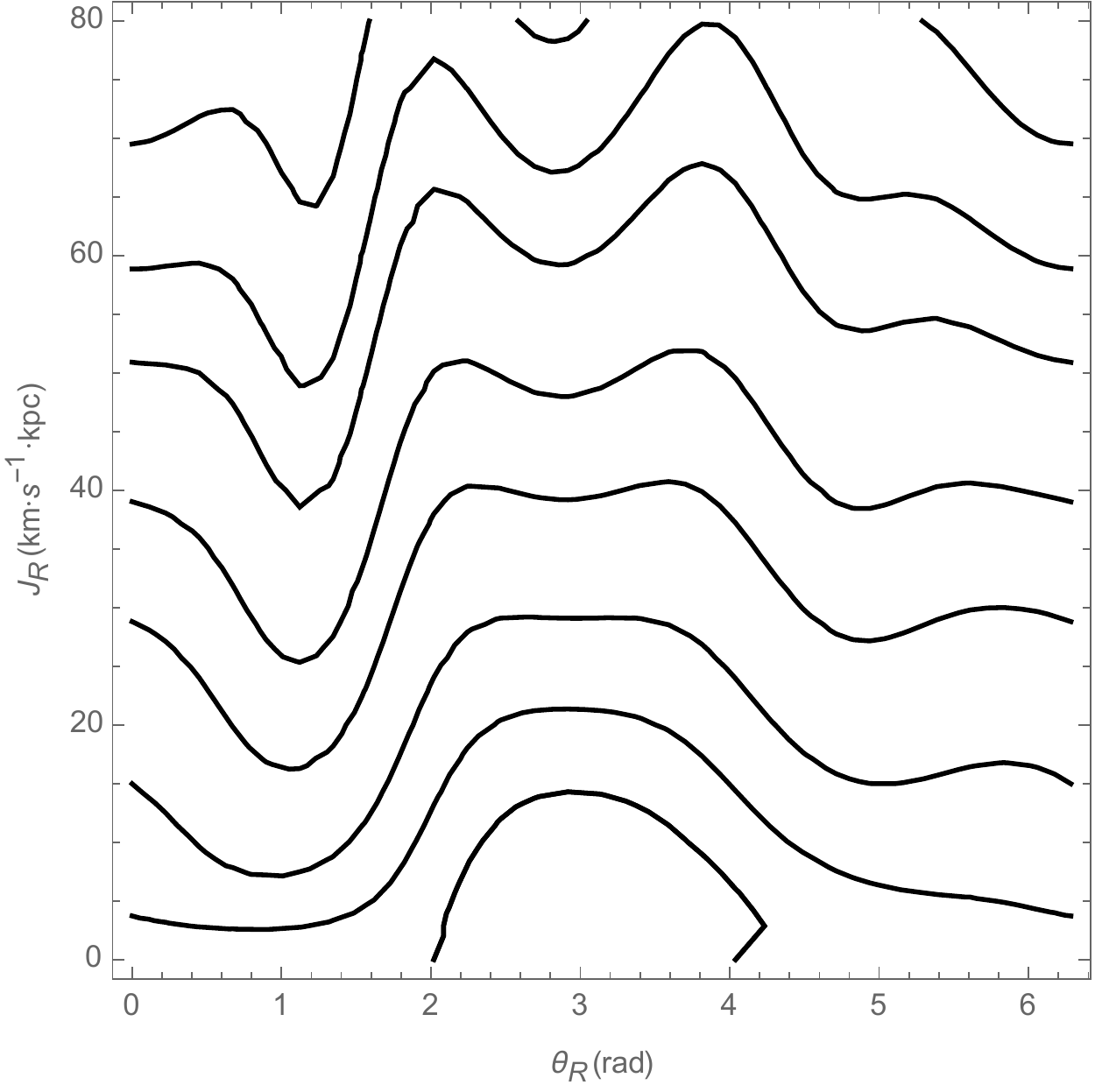}
  \caption{Isocontours of the distribution functions in the $(\theta_R,J_R)$ space at the
    point $(R,\phi,z)=(8\Kpc,0,0)$ of the Galactic plane. Top panel:
    $f_0(\theta_R,J_R)$. Bottom panel: $f(\theta_R,J_R)$. The contours
    enclose (from bottom to top) 12, 21, 33, 50, 68, 80, 90, 95, and
    99\% of the stars.}\label{fig:DF0}
\end{figure}
\begin{figure}
  \centering
  \includegraphics[width=\columnwidth]{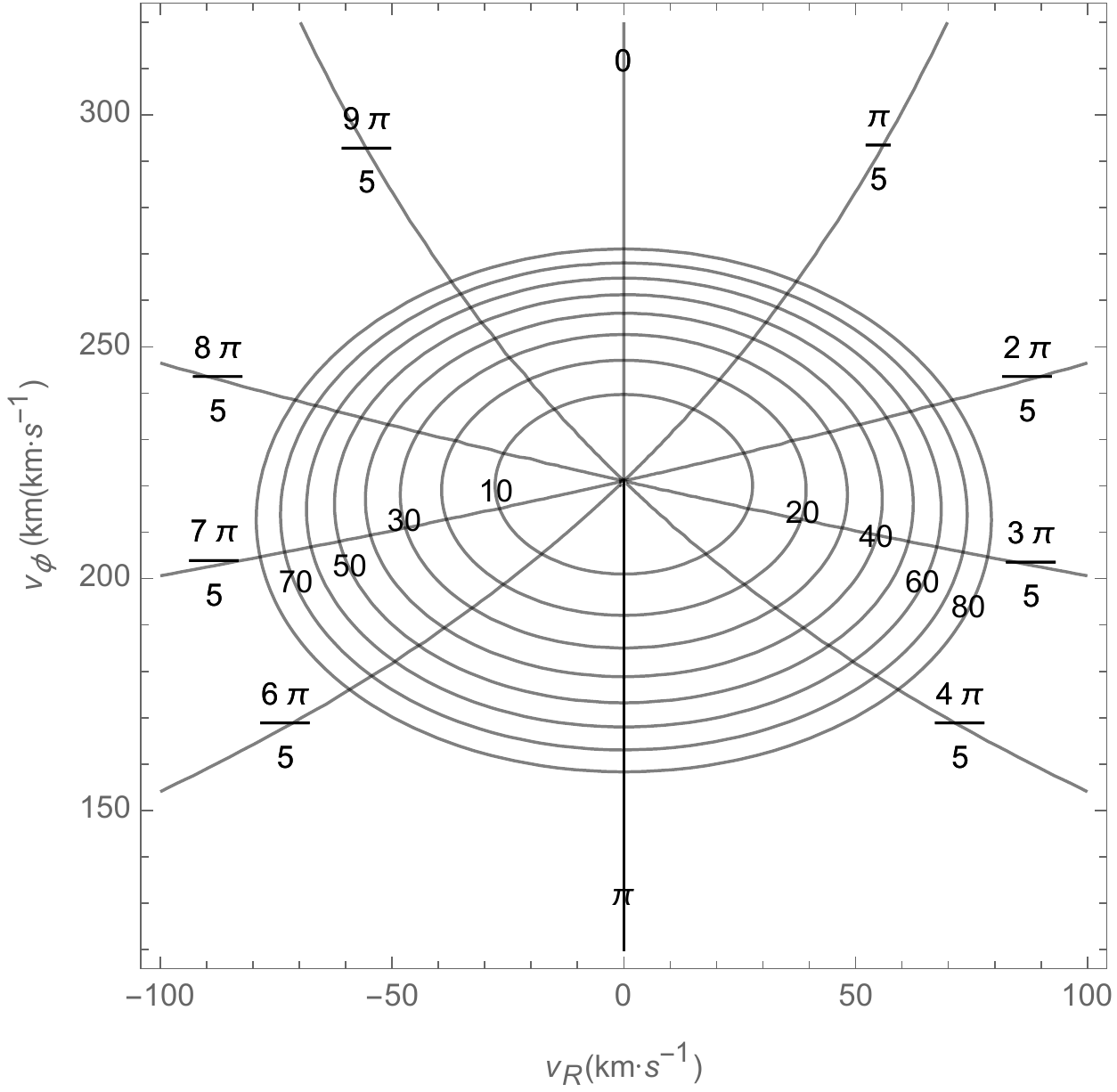}
  \caption{Curves of constant $\theta_R$ and $J_R$ in velocity space at
    $(R,\phi,z)=(8\Kpc,0,0)$ for the
    \protect\cite{BT2008} Model~I potential. See also \citet{McMillan2011}.}\label{fig:map}
\end{figure}
\begin{figure}
  \centering
  \includegraphics[width=\columnwidth]{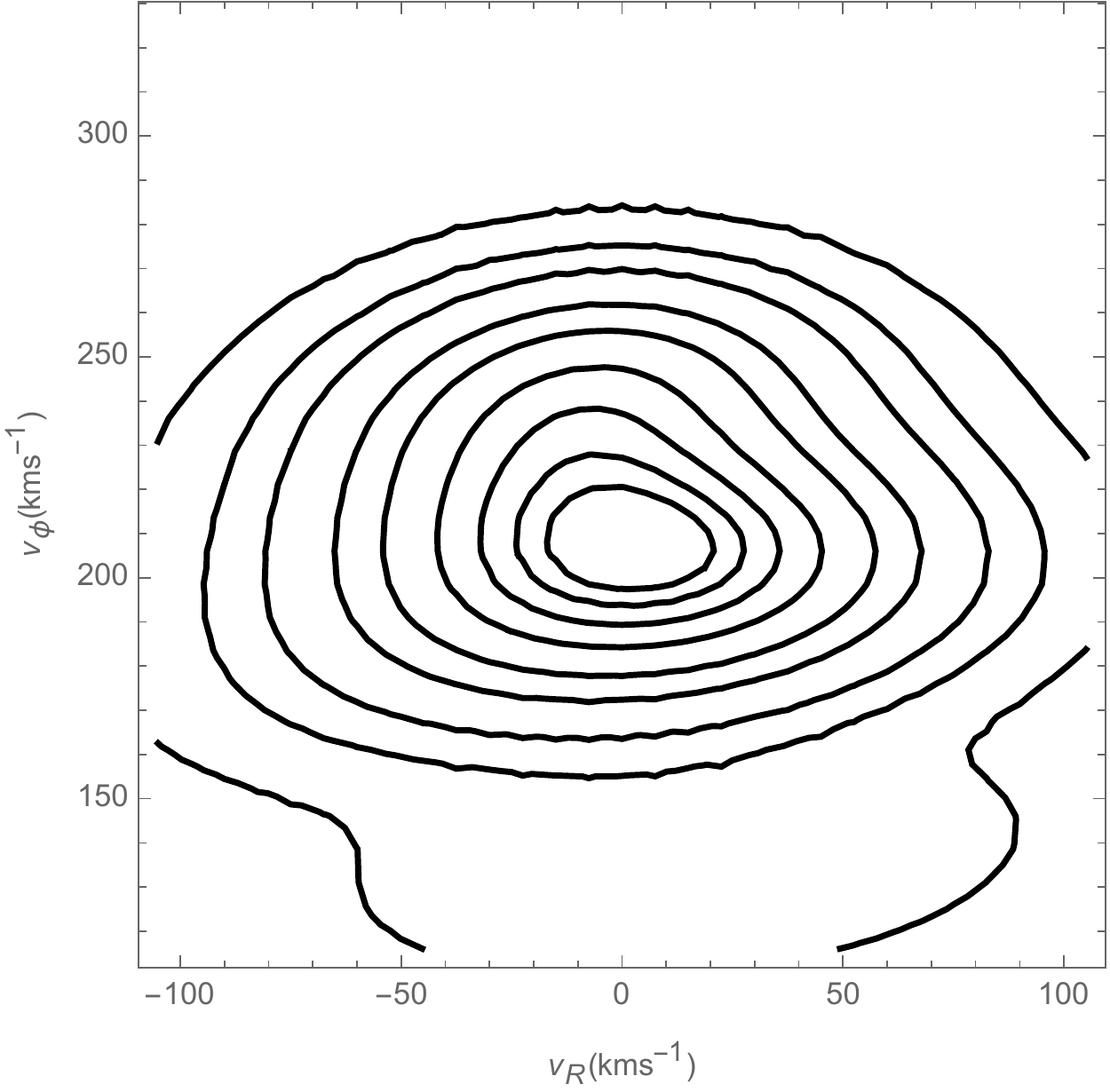}
  \includegraphics[width=\columnwidth]{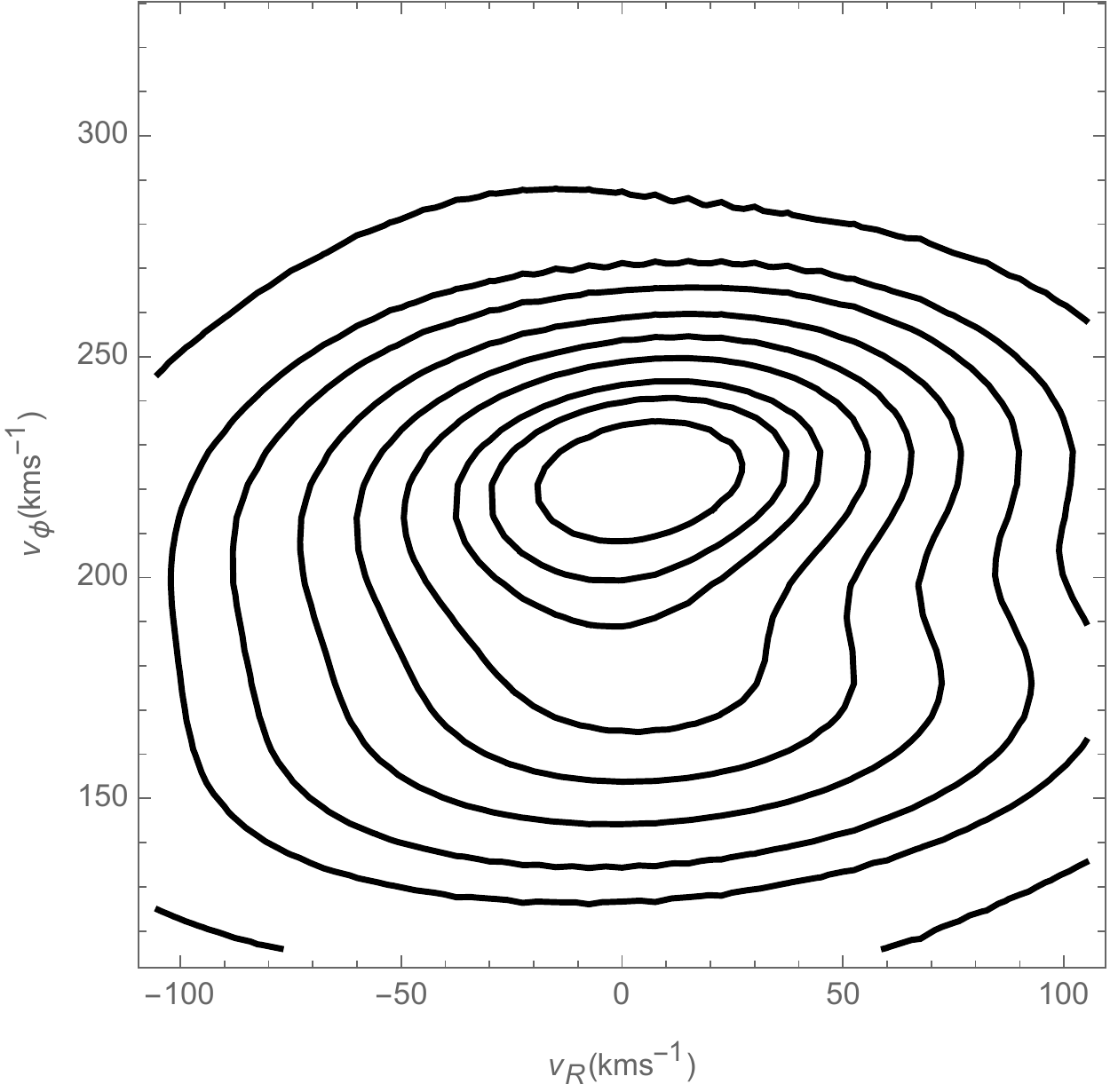}
  \caption{Isocontours of the velocity distribution functions $f(v_R,v_\phi)$ at two
    points of the Galactic plane. Top panel:
    $(R,\phi,z)=(8\Kpc,0,0)$. Bottom panel:
    $(R,\phi,z)=(6\Kpc,0,0)$. The contours enclose (from the inner to
    the outer) 12, 21, 33, 50, 68, 80, 90, 95, and 99\% of the
    stars.}\label{fig:DF1}
\end{figure}
\begin{figure}
  \centering
  \includegraphics[width=\columnwidth]{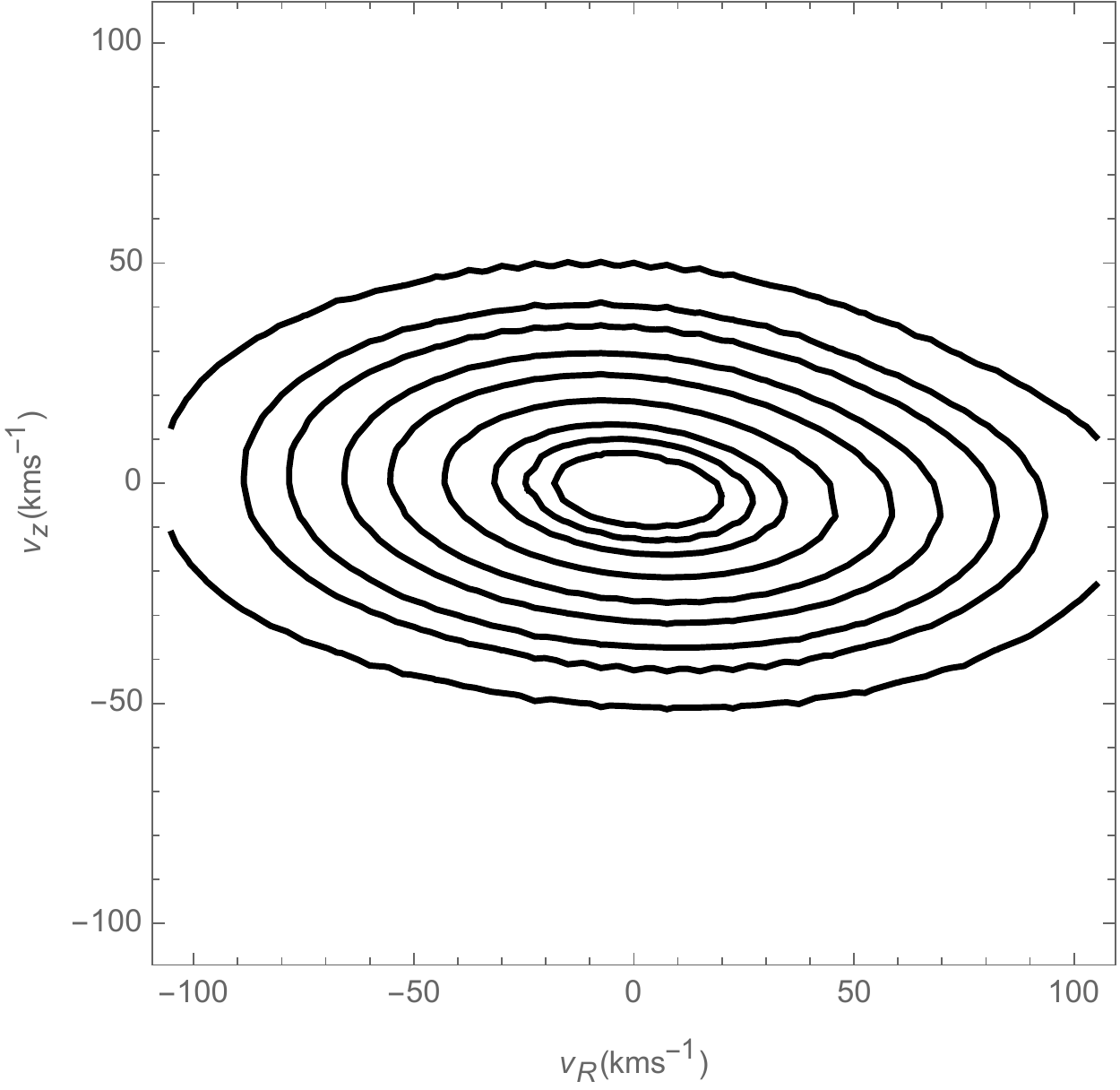}
  \includegraphics[width=\columnwidth]{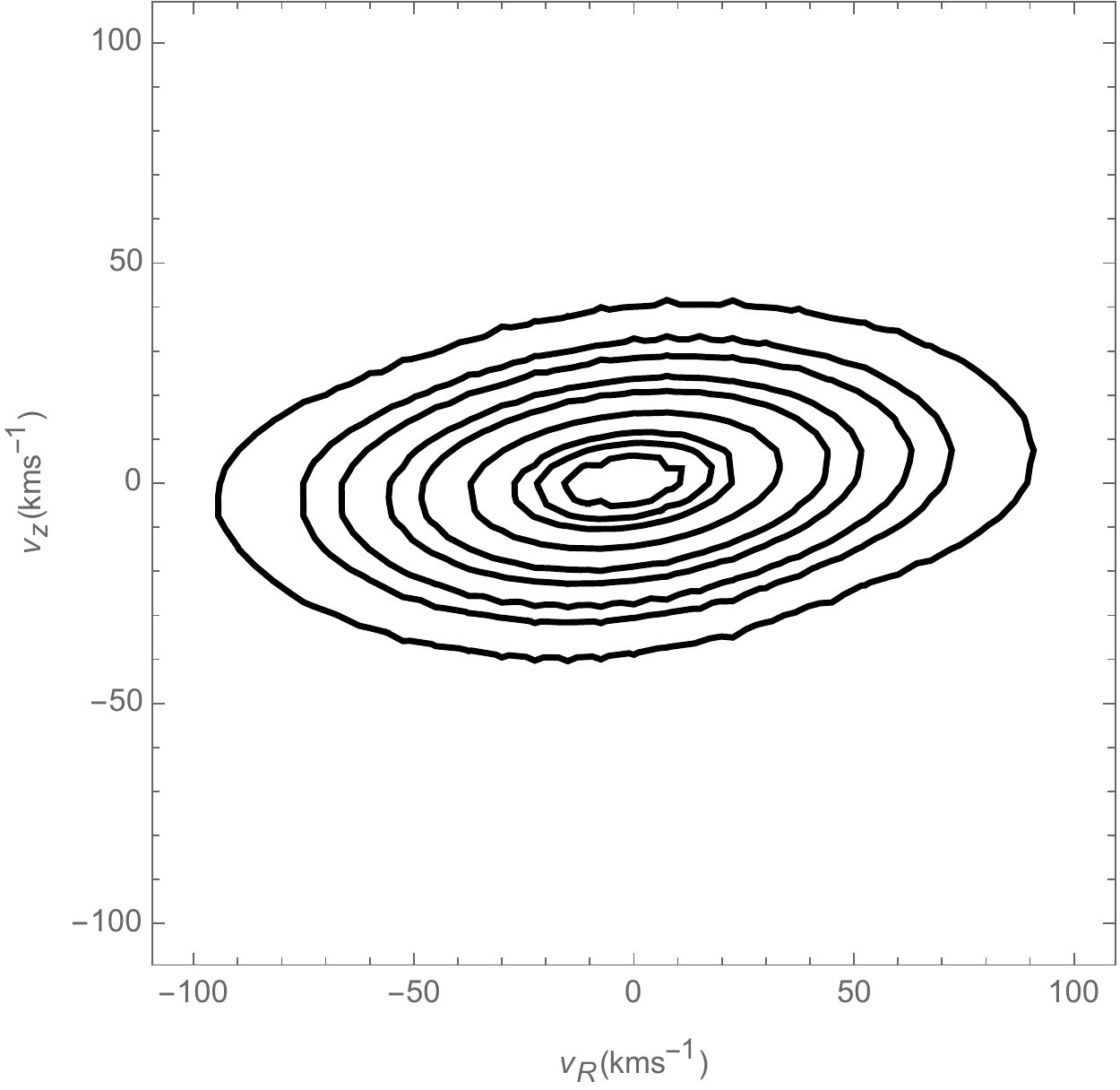}
  \caption{Isocontours of the velocity distribution functions $f(v_R,v_z)$ at two points
    at $z=0.3\Kpc$ from the Galactic plane. Top panel:
    $(R,\phi,z)=(7.5\Kpc,0,0.3\Kpc)$. Bottom panel:
    $(R,\phi,z)=(9.5\Kpc,0,0.3\Kpc)$. The contours enclose (from the
    inner to the outer) 12, 21, 33, 50, 68, 80, 90, 95, and 99\% of the
    stars.}\label{fig:DF2}
\end{figure}
Our computation of the exact form of the perturbed DF away from the
main resonances also allows us to study the detailed behaviour of $f =
f_0 + \epsilon f_1$ at a given point in configuration space
$(R,\phi,z)$, in terms of the actions and angles, and compare it with
the unperturbed version of the DF, $f_0$. First, let us note that the
dimensions of phase-space, given the constraint of a fixed point in
configuration space, $(R,\phi,z)=\mathrm{constant}$, decrease from 6
to 3, even when the DF depends both on actions {\it and} angles. We
focus on the case $(R,\phi,z)=(8\Kpc,0,0)$ (i.e, the typical position
of the Sun in our model) and we add the constraint $J_z=0$,
additionally decreasing the dimensionality of phase-space to 2
dimensions.

The two variables that we choose to display are $(\theta_R,J_R)$. The
other angles and actions are constrained by
$R=\Rg(J_\phi)-\sqrt{2J_R/\kappa(J_\phi)}\cos\theta_R$, $J_z=0$,
$\theta_\phi=\phi+\Delta\phi(J_\phi,J_R,\theta_R)$, and
$\theta_z=\upi/2$ (because $z=0$). In practice we solve numerically
the first of this constraints for each pair $(\theta_R,J_R)$ to get
$J_\phi$, and it is then trivial to get $\theta_\phi$. In the case of
the unperturbed DF, $f_0$, the true dependence is obviously on $J_R$
and $J_\phi$, but we can translate it in terms of $(\theta_R,J_R)$ in
terms of the above constraints at a fixed point in configuration
space.

The comparison between $f_0(\theta_R,J_R)$ and $f(\theta_R,J_R)$ is
shown in \Fig{fig:DF0}. As is apparent from this figure, both $f_0$
and $f$ decrease with $J_R$, but $f_0$ is symmetric with respect to
$\theta_R=\upi$ while $f$ not, which is due to the
$\exp(\pm\img\theta_R)$ terms. The asymmetries in \Fig{fig:DF0} can be
directly translated to features in the $(v_R,v_\phi)$ velocity
space. To visualize this transformation, we are helped by the map in
\Fig{fig:map}, which shows how to associate $(\theta_R,J_R)$ with
$(v_R,v_\phi)$ at $(R,\phi,z)=(8\Kpc,0,0)$. This figure displays
curves of constant $J_R$ and $\theta_R$ in the $(v_R,v_\phi)$ space
\citep[see also][]{McMillan2011}. Each of the central circular curves
represents a value of $J_R$, while the lines that radiate from
$(v_R,v_\phi)=(0,220\kmsec)$ represent constant values of $\theta_R$.

It is then interesting to check the behaviour of the velocity
distribution function at such a particular point in space, i.e.,
$f(v_R,v_\phi,v_z)$ at constant $(R,\phi,z)$, as we can e.g., compare it
to the velocity of the stars in the Solar neighbourhood (the small
volume around the Sun where, to date, detailed enough kinematic data
are present). As a matter of fact, velocity-space substructures in the
Solar neighbourhood, called {\it moving groups}, have observationally
been shown to be composed of stars of different ages and chemical
compositions \citep[e.g.][]{Dehnen1998, Chereul1999, Famaey2005,
  Famaey2008, Pompeia2011}. For this reason, they are most likely
associated to perturbations from the bar and spiral arms
\citep[e.g.,][]{Dehnen2000,Minchev2010,Antoja2011,Quillen2011}. Our
model is based on a single spiral perturber, and is valid only away
from the main resonances, so we do not expect the model to reproduce
all the observed features. Nevertheless, it is interesting to look at
the trend (note that the ``Solar neighbourhood" in our model is indeed
away from the main resonances as we chose parameters such that ${\rm
  ILR}=1.56\Kpc$ and ${\rm CR}=11.49\Kpc$).

In \Fig{fig:DF1} (top panel) we show the perturbed DF $f(v_R,v_\phi)$
at $(R,\phi,z)=(8\Kpc,0,0)$. We see how the effect of the perturbation
is to deform the density contours so that the stars are not anymore
distributed symmetrically between positive and negative $v_R$. In
particular there is an excess of stars slightly lagging rotation and
moving outwards around $v_R \simeq 30\kmsec$ and
$v_\phi=210\kmsec$. This particular configuration of the density
contours is reminiscent of that created by the Hyades moving group in
the Solar neighbourhood\footnote{However, it is likely that the Hyades
  moving group is a resonant feature
  \citep{Sellwood2010,Hahn2011,McMillan2011,McMillan2013}, thereby not
  precisely reproduced by the present model.}. These features can be
easily interpreted in light of \Fig{fig:DF0} and \Fig{fig:map}. For
example, fixing $J_R=30\kmsec\Kpc$ and moving clockwise from
$\theta_R=0$, we first encounter in \Fig{fig:DF0} an underdensity at
$\theta_R\approx \upi/5$. Then the density increases again at
$\theta_R=\upi/2$, forming in \Fig{fig:DF1} (top) the Hyades-like
distortion. At $\theta_R\approx\upi$ it is almost constant, to
slightly decrease again for $\theta_R>\upi$. The general velocity
distribution is slightly skewed towards negative radial velocities. In
the bottom panel of \Fig{fig:DF1}, we then also show $f(v_R,v_\phi)$
at $(R,\phi,z)=(6\Kpc,0,0)$. Here we find more stars that in the
previous case at $v_\phi<220\kmsec$. Moreover, the two configurations
in the DFs of \Fig{fig:DF1} explain why there is a net $\avvR<0$
motion at $(R,\phi,z)=(8\Kpc,0,0)$ in the Galaxy, while $\avvR>0$ at
$(R,\phi)=(6\Kpc,0,0)$, due to the asymmetry of the general velocity
distribution.

In \Fig{fig:DF2} we now show $f(v_R,v_z)$ fixing $v_\phi=\vc(7.5\Kpc)$
and $(R,\phi,z)=(7.5\Kpc,0,0.3\Kpc)$ (top panel) and
$v_\phi=\vc(9.5\Kpc)$ and $(R,\phi,z)=(9.5\Kpc,0,0.3\Kpc)$ (bottom
panel), hence at $z=0.3\Kpc$ height from the Galactic
plane\footnote{To obtain the DFs at $z=-0.3\Kpc$ it is sufficient to
  flip $v_z$ with $-v_z$.}. The former case has $\dvz<0$ and
$\avvR>0$, while the latter $\dvz>0$ and $\avvR<0$. The consequence of
the perturbation is a tilt of the velocity ellipsoid in the $v_R-v_z$
space, that has opposite sign in the two points. Such a tilt would be
impossible, by construction, with the unperturbed $f_0$ distribution
function which is plane-parallel, and has a similar amplitude of
that found by studies of stars in the Solar neighbourhood
\citep[e.g.,][]{Pasetto2012}. The velocity ellipsoid is thus clearly
influenced by the spiral potential, and this intuitively explains why
there is a transition from positive to negative mean vertical motion
precisely in between the arms and in the middle of the arms (where the
mean radial motion is maximal), because the ellipsoid becomes
plane-parallel again. Nevertheless, a tilt of the ellipsoid alone
cannot cause a net vertical motion, as the average $v_z$ would still
be $0$. But this tilt is actually accompanied by a lopsidedness of the
$v_z$ distribution, which is maximal when the tilt is maximal.

\section{Conclusion and perspectives}\label{sect:conclusions}
This work presents a general way to calculate the effects of a
non-axisymmetric gravitational disturbance on an axisymmetric DF,
$f_0$, describing the phase-space density of stars in a collisionless
stellar system (i.e., governed by the collisionless Boltzmann
equation). We assume that the axisymmetric $f_0$ alone solves the
collisionless Boltzmann equation in an axisymmetric potential $\Phi_0$
where the relationship between the ordinary positions and velocities
and the action and angle variables are known (\Sec{sect:DF}).

We apply this method to construct a 3D model of the Milky Way's thin
disc, where the non-axisymmetric gravitational disturbance $\epsilon
\Phi_1$ is a Fourier mode in azimuth (\Sec{sect:lCBE}). In particular,
we chose to describe bisymmetric spiral arms with a $\sim {\rm
  sech}^2$ vertical falloff (\Sec{sect:results_mod}). As a result, we
obtain formulas for the DF and its zeroth and first order moments
(density and mean motions) that are shown to be in agreement with a
numerical test-particle simulation representing the effect of the same
bisymmetric spiral arms on the Milky Way's thin disc
(\Sec{sect:results_mom}). In particular, we estimate for the first
time the reduction factor for the vertical bulk motions of a stellar
population compared to the case of a cold fluid.

An inspection of the DF at given points in 3D configuration space
(\Sec{sect:results_DF}) also helps to interpret these macroscopic
properties of the stellar system. One interesting result is that the
spiral arms induce a tilt and a lopsidedness in the $v_R-v_z$ velocity
ellipsoid that changes of sign and magnitude as a function of the
position of the point where it is calculated w.r.t. the spiral
arms. In addition, it is shown that distortions typical of moving
groups such as the Hyades are naturally reproduced in velocity space.

We nevertheless point out that our results here are only valid away
from the main resonances. Indeed, our method consists in a linear
treatment of the collisionless Boltzmann equation, i.e., it assumes
that the non axisymmetric gravitational disturbance $\epsilon\Phi_1$
and DF response $\epsilon f_1$ are small. In particular, $f_0$ should
always be larger than $\epsilon f_1$ in order to preserve physical
meaning. While most of the non-axisymmetric gravitational disturbances
of the Milky Way are indeed much smaller than its background
axisymmetric gravitational potential, certain regions of phase-space
are particularly affected by the perturbations. These are the
resonances, where the rotational, radial, and vertical frequencies and
the perturbation pattern speed are commensurable. The linear regime
breaks down at the resonances, as is evident from \Eq{eq:F}, whose
denominator vanish at the resonances. Even if there is an infinite
number of resonances, those that affect a significant portion of
phase-space are rare. In our treatment they appear for example at the
corotation and Lindblad resonances that, in the case of the spiral
arms we chose in this paper, are all quite far from the Solar
neighbourhood. The same cannot be stated in the case of the bar, where
the outer Lindblad resonance is probably close to the Sun.  One way to
treat the resonances that we will explore in forthcoming work is to
pass, in their vicinity, to another system of angle-action variables
(fast and slow variables), that allows to focus on the librations
around the resonant orbits, neglecting all the high frequency motions
\citep[see][]{BT2008}.

Another future issue, even more complex to treat, is related to the
non-linear effects due to the presence of more than one perturber. In
the linear perturbation theory presented here, the effect of more than
one perturber would simply be the linear combination of the single
responses. However, from numerical studies (Monari et al. in prep.),
it can be shown that non-linear effects arise simply by superposing
different perturbers, as the bar and spiral arms. This is especially
important in terms of the amplitude of the vertical breathing mode
generated by the spirals in this work, which is qualitatively similar
to observations \citep{Williams2013}, but not quantitatively. The
effect of multiple perturbers could be especially important in that
case. Future analytic calculations should investigate this
question. Also, in the present work, we concentrated on the response
of a given disc stellar population in equilibrium to a perturbing
three-dimensional spiral potential, but we did not investigate yet the
conditions for self-consistency, which, especially in 3D, is a more
complex problem than the present one, to be treated in the future too.

Finally, we note that, while we used the adiabatic and epicyclic
approximations to estimate the angle and action variables in this
work, the method to obtain the distribution function that we present
at the beginning of the paper is completely general
(\Sec{sect:lCBE_gen}). Our choice of using a Schwarzschild
distribution function to represent the axisymmetric equilibrium
configuration can trivially be generalized to other forms of the
distribution function. Moreover, our results can also in principle be
used with more sophisticated approximations of the angles and actions
in the Milky Way potential already present in the literature. For this
reason, the method presented in this paper will be helpful in the
future to dynamically characterize the Milky Way disc stellar
kinematic information that will be provided by upcoming large
astrometric and spectroscopic surveys of the Galaxy, as it offers the
possibility to interpret the latter in the dynamical sense (rather
than just subtracting the residuals from a fiducial axisymmetric
model), using a rather low number of free parameters.

\section*{Acknowledgements}
This work has been supported by a postdoctoral grant from the {\it
  Centre National d'Etudes Spatiales} (CNES) for GM.

\bibliographystyle{mn2e}
\bibliography{mn-jour,f1pertbib}

\begin{appendix}

\end{appendix}

\label{lastpage}

\end{document}